\documentclass[journal]{IEEEtran}
\usepackage[tight,footnotesize]{subfigure}

\usepackage{algorithmic}
\usepackage[linesnumbered,ruled]{algorithm2e}
\usepackage{graphics,epstopdf}
\usepackage{balance}
\usepackage{listings}
\usepackage{comment}
\usepackage{enumitem}
\usepackage{xcolor}
\usepackage{fixmath}
\usepackage{cite}
\usepackage{amsmath,amssymb,amsfonts}
\usepackage{graphicx}
\usepackage{textcomp}
\usepackage{tabularx}
\usepackage{booktabs}
\usepackage{longtable}
\usepackage{multirow}
\usepackage{colortbl}
\usepackage{hhline}
\usepackage{tabularx,colortbl}
\usepackage{stmaryrd}
\usepackage{hyperref} 
\usepackage{cite}
\usepackage{enumitem}
\usepackage{soul}
\usepackage{bm} 
\usepackage{mathtools} 
\usepackage{mathrsfs} 
\usepackage{multirow}
\usepackage{array}
\usepackage[ruled,linesnumbered]{algorithm2e}
\definecolor{seagreen}{rgb}{0.18, 0.55, 0.24}
\usepackage{caption}

\graphicspath{ {images/} }

\begin{document}

\title {Detection of Lying Electrical Vehicles in Charging Coordination Application Using Deep Learning}

\author{\IEEEauthorblockN{
  Ahmed Shafee,
  Mostafa M. Fouda,~\IEEEmembership{Senior~Member,~IEEE,}
  Mohamed Mahmoud,~\IEEEmembership{Member,~IEEE,}\\
  Waleed Alasmary,~\IEEEmembership{Senior~Member,~IEEE,} Abdulah J. Aljohani,
  and
  Fathi Amsaad,~\IEEEmembership{Member,~IEEE}}
\thanks{Corresponding author: Mostafa Fouda.}
\thanks{A. Shafee, and M. Mahmoud are with the Department of Electrical and Computer Engineering, Tennessee Tech. University, Cookeville, TN 38505 USA (e-mail: aashafee42@students.tntech.edu; mmahmoud@tntech.edu).}
\thanks{M. Fouda is with the Department of Electrical and Computer Engineering, Tennessee Tech. University, Cookeville, TN 38505 USA, and the Department of Electrical Engineering, Faculty of Engineering at Shoubra, Benha University, Egypt (e-mail: mfouda@ieee.org).}
\thanks{W. Alasmary is with the Department of Computer Engineering, Umm Al-Qura University, Makkah, Saudi Arabia.}
\thanks{A. Aljohani is with the Department of Electrical and Computer Engineering, King Abdulaziz University, Jeddah 21589, Saudi Arabia, and also with the Center of Excellence in Intelligent Engineering Systems (CEIES),
King Abdulaziz University, Jeddah 21589, Saudi Arabia.}
\thanks{F. Amsaad is with the School of Information Security and Applied Computing (SISAC), Eastern Michigan University (EMU), MI 48197 USA.}
}
\maketitle
\IEEEpeerreviewmaketitle

\begin{abstract}
The simultaneous charging of many electric vehicles (EVs) stresses the distribution system and may cause grid instability in severe cases. The best way to avoid this problem is by charging coordination. The idea is that the EVs should report data (such as state-of-charge (SoC) of the battery) to run a mechanism to prioritize the charging requests and select the EVs that should charge during this time slot and defer other requests to future time slots. However, EVs may lie and send false data to receive high charging priority illegally. In this paper, we first study this attack to evaluate the gains of the lying EVs and how their behavior impacts the honest EVs and the performance of charging coordination mechanism. Our evaluations indicate that lying EVs have a greater chance to get charged comparing to honest EVs and they degrade the performance of the charging coordination mechanism. Then, an anomaly based detector that is using deep neural networks (DNN) is devised to identify the lying EVs. To do that, we first create an honest dataset for charging coordination application using real driving traces and information revealed by EV manufacturers, and then we also propose a number of attacks to create malicious data. We trained and evaluated two models, which are the multi-layer perceptron (MLP) and the gated recurrent unit (GRU) using this dataset and the GRU detector gives better results. Our evaluations indicate that our detector can detect lying EVs with high accuracy and low false positive rate.
\end{abstract}


\section{Introduction}
\label{Introduction}
  Recently, there has been a growing interest to adopt more green transportation in an effort to reduce the associated carbon emissions and reduce the dependency on crude oil. 
  As a result, a promising direction is to replace gasoline-powered vehicles by electric vehicles (EVs). In recent years, the number of EVs traveling on the roads has been experiencing a dramatic increase \cite{inproceedings}.
 Another important aspect that helps in the wide use of EVs is the recent trend of adopting more renewable energy sources at consumer side by installing solar cells on the roof of houses. Due to the intermittent nature of these sources, EVs can be used to store the excess energy. Regardless of the numerous advantages that are introduced by the EVs, several challenges need to be addressed when dealing with deploying a large number of EVs in the street.
 
 The uncoordinated and simultaneous charging of EVs may stress the power system and instablize the grid in severe cases \cite{6863680, 5664815}.
  Accordingly, charging coordination mechanisms should be used to avoid this problem by balancing charging demand and power supply \cite{Schuller2015}. The idea is that each EV should report data (such as state of charge (SoC) of the battery) to a charging coordinator (CC) that uses these data to prioritize the EVs' charging requests and allow the high-priority EVs (typically the ones that have low SoC) to charge and defer other requests to future time slots \cite{8946265}. Many papers in the literature have developed charging coordination mechanisms \cite{5986769, 6062663, Tushar2014PrioritizingCI}. However, \textit{these mechanisms assume that EVs report correct data}, but such assumption can not be guaranteed because EVs are motivated to report false data (e.g. lower SoC) so that they can charge quickly without deferral. 
  
  In this paper, we first evaluate the impact of lying EVs on the charging coordination application. Specifically, we focus on evaluating the gains the lying EVs achieve, the harm they cause to honest EVs and also the impact of the lying EVs on the performance of the charging coordination mechanism. Our evaluations confirm that reporting false information is beneficial for the EVs as they charge first, honest EVs are harmed by not charging or charging late, and the performance of the charging coordination mechanism degrades because by reporting low SoC falsely, excess power is assigned to the lying EVs which may result in unusing all the power available for charging.
  
  In order to resolve this issue, we devise a machine learning model to detect the lying EVs. We first create a dataset for the charging coordination application. To do that, we use the real driving traces of vehicles provided in \cite{TaxiCabs} and real EVs information provided by Kia automotive manufacturer in \cite{kiaSoul}. Then, we propose several attacks for reporting false SoC by EVs to the charging coordination mechanism. To balance the data, we use adaptive synthetic sampling approach (ADASYN) \cite{ADASYN} to compensate the samples of the minor class to be equal, or nearly equal, to the samples of the major class. The ADASYN balances the dataset by computing the ratio between the honest dataset (minor class) and the malicious dataset (major class), and producing synthetically more honest data samples for the compensation process. Finally, we use the dataset to create deep learning models to detect lying EVs. Deep learning has been used in many applications because of its high accuracy \cite{Shokri_ML2015} such as face recognition, intrusion detection, and speech analysis.
  In this paper, multi-layer perceptron (MLP) network \cite{MlpArticle} and deep recurrent neural network (RNN) \cite{5947611} are the machine learning models that are selected for handling the time series nature of our dataset. However, choosing the best hyper-parameter set for the deep neural networks is considered a hard optimization problem. As a result, we use a version of a genetic optimization \cite{797971} technique to find the best architectures for our detectors.
  To evaluate our detectors, extensive experiments have been conducted. Our evaluations confirm that our detectors can identify the lying EVs with very high accuracy and low false positive rate.

Comparing to the literature, this paper makes the following main contributions:
\begin{enumerate}
    \item We investigate the impacts of reporting false data by EVs to charging coordination mechanism. 
    \item We create a new dataset for charging coordination application using real vehicles routes and information reported by an automotive company. We also propose several attacks to report false data. Although, our focus is to create a dataset that is used to train detectors to identify lying EVs, our dataset can also be used for other applications such as prediction of energy demand.
    \item We propose deep-learning models to identify lying EVs and run extensive experiments to evaluate the models.
\end{enumerate}

The remainder of this paper is organized as follows. The related work is discussed in section \ref{Related Work}. Section \ref{System Model} describes the system model.
The evaluation of the impact of lying EVs on charging coordination mechanisms is presented in section \ref{Problem Formulation}.  Preliminaries are presented in section \ref{Preliminaries}. The dataset created for pure EVs is presented in section \ref{Dataset}. The proposed deep learning based detectors are discussed in details in section \ref{Design of Electric vehicle detector}. Our evaluations to the proposed detectors and experimental results are discussed in section~\ref{exp}.  Our conclusions are presented in section~\ref{Conclusion}.

\section{Related Work} 
\label{Related Work}
In this section, we first present the existing works on charging coordination mechanisms. Then, we discuss the existing datasets for EVs. 
\subsection{Charging Coordination Mechanisms}
Several works have investigated charging coordination mechanisms for EVs in the literature. Arias et al. \cite{optimization1} have proposed an optimized charging coordination mechanism for finding the best charging schedule that meets the EVs requirements and the operational capacity of the electrical distribution system. The authors use three optimization algorithms including the tabu search (TS), the greedy randomized adaptive search procedure (GRASP) and a new hybrid optimization algorithm that uses both TS and GRASP. 
Hajforoosh et al. \cite{RealTimeCharging} have proposed a charging coordination mechanism based on the fuzzy discrete particle swarm optimization (FDPSO) algorithm and the fuzzy genetic algorithm (FGA). The main objective of the proposed mechanism is to maximize the amount of electrical power delivered to the EVs and minimize the grid losses, distribution transformer loading and the cost associated with the energy generation. 
Franco et al. \cite{7031457} have proposed a charging coordination mechanism for unbalanced electrical distribution systems using a mixed integer linear programming (MILP) technique. The proposed mechanism aims at reducing the energy cost of EV charging taking into consideration several factors such as the three phase circuit's representation and the load imbalance producing non-linear programming model that could be converted into a linear model using linearization techniques that could be solved easily. However, the above works do not consider preserving the privacy of the EVs' drivers.

Mahmoud et al. \cite{7564967} have introduced a privacy-aware charging coordination mechanism that preserves the privacy of the EVs drivers while still optimizing the power supplied to the EVs without exceeding the total charging capacity. The idea is that each EV encrypts its charging request in addition to a one-time secret key with the CC's public key so that only the CC can decrypt the request. Then, the CC uses the data provided to compute charging schedules and encrypt them with the secret keys sent by the EVs. At the same time, random noise is added to the data sent by EVs to prevent linking the charging requests sent from one EV to preserve privacy. In addition, a modified version of knapsack optimization algorithm is used to schedule charging of EVs.  

Baza et al. \cite{8946265} have introduced a decentralized charging coordination mechanism using the blockchain and smart contract technology. 
Pazos-Revilla et al. \cite{DBLP:journals/corr/abs-1905-04666} have proposed two privacy-preserving charging coordination mechanisms. The first mechanism has a centralized architecture, and uses blind signature cryptosystem for anonymous authentication. The second mechanism has a decentralized architecture where various EVs run the mechanism in a distributed manner. The idea is that an EV is selected to act as a head node that decrypts the EVs' aggregated charging demand and broadcasts the aggregated demand to the EVs  without being able to learn the individual data of the EVs to preserve privacy. Then, each EV can compute its charging schedule without exceeding the maximum charging capacity.

To the best of our knowledge, \textit{the existing works in the literature assume that EVs report correct data} to the charging coordination mechanism. This assumption can not be guaranteed because the EVs are \textit{motivated to report false data to charge quickly} without deferral. In addition, none of the existing works have studied the impact of reporting false data on the charging coordination mechanism or proposed a scheme to identify the EVs that report wrong data.

\subsection{Dataset for EVs}
Khavan-Hejazi et al. \cite{Developing_a_Test_DataSet} have created a dataset for hybrid EVs that contains a minute-by-minute SoC data as well as the nodal and temporal charging load of plug-in hybrid EVs (PHEVs). The dataset has been created by using the driving traces of 536 gasoline-powered taxi vehicles equipped with global positioning system (GPS) in San Francisco and the nominal operation data of various PHEVs brands. The main goal is to find the SoC and charging patterns of the vehicles with the same movement patterns but with various PHEV technologies. 
As the vehicles are moving, they switch between electrical energy and gasoline. The dataset considers two types of PHEVs. The first one is the charge depleting type that switches to the gasoline after depleting of the electric battery. The second type is the charge blending type of PHEV which switches to the gasoline engine to increase the torque of the vehicle even if the electric battery is not depleted. 

Qinglong in \cite{DBLP:journals/corr/abs-1802-04931} have proposed a model to predict the amount of stored energy in the EVs of a specific area. A neural network model combined with linear chain conditional random field (CRF) are used to create the model. The dataset which is used for training the neural network is created by using the GPS trajectories of a set of taxis in Beijing city. The paper assumes that the SoC of each EV changes linearly with respect to the distance driven by the vehicle. 
However, the linear model used for creating the EV's dataset is not mentioned. The paper also assumes that all the EVs recharge to the full capacity at the begining of each day, which may not be guaranteed practically. Moreover, the dataset has not been shared publicly.

Oh et al. \cite{Oh2019VehicleED} have provided a dataset for vehicles including gasoline powered vehicles, hybrid EVs (HEVs), PHEVs, and EVs. The dataset is collected using a real-world driving scenarios for a set of vehicles including driving in different environments from highways to traffic-dense regions. The information in this dataset is divided into static and dynamic. The static information contains the vehicle's parameters such as the vehicle type and engine configuration. The dynamic information contains time-stamped driving records of the vehicle such as the vehicle's speed, and battery current, voltage and SoC.
However, this dataset contains the data of only three EVs, which is very small dataset for training machine learning models. 

Unlike these works, in this paper, we create \textit{a new dataset for charging coordination application including data for lying EVs.}

\section{Network and Threat models}
\label{System Model}
\begin{figure}[!t]
\centering
\captionsetup{justification=centering}
\includegraphics[width=\columnwidth]{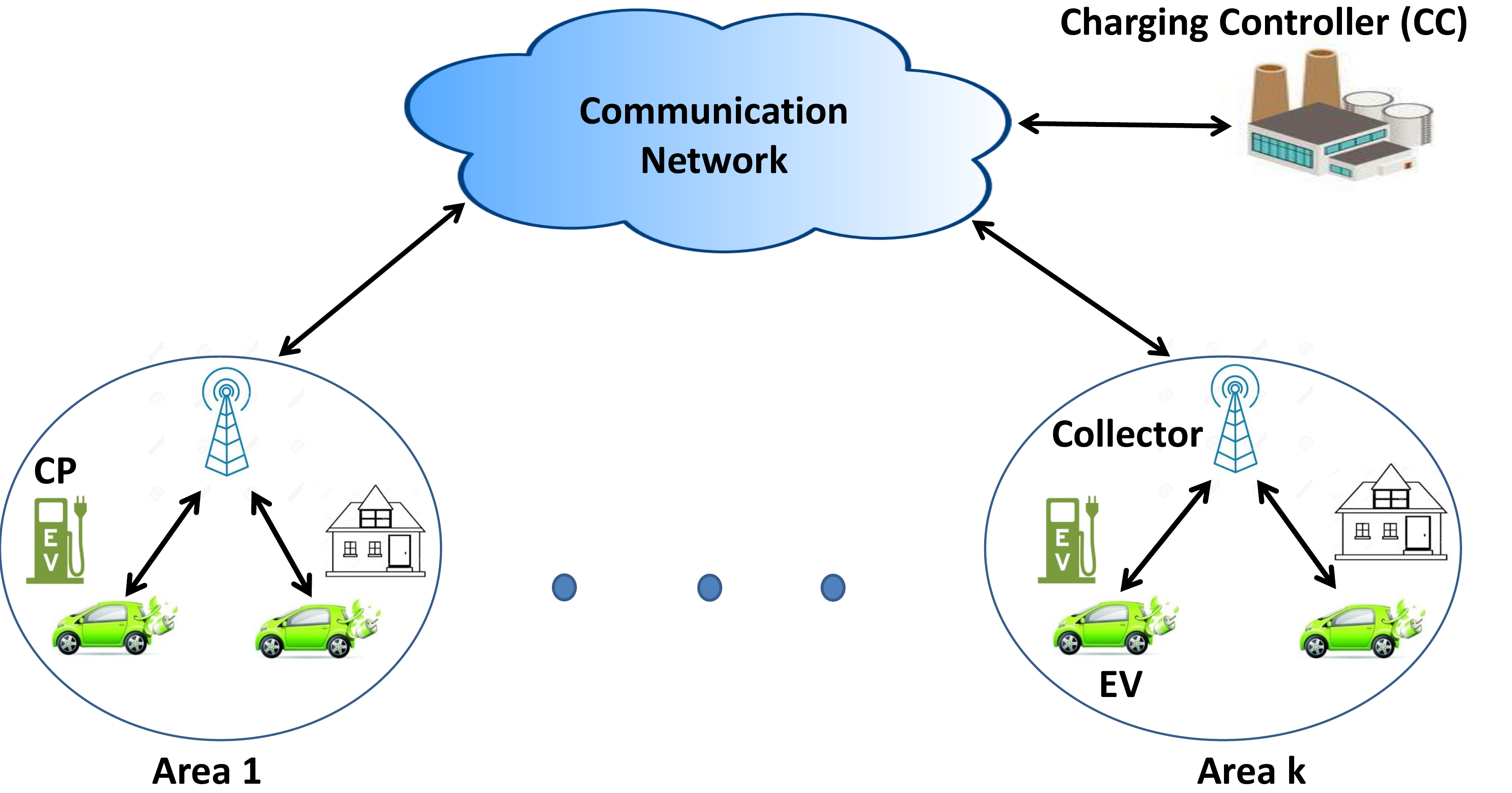}
\caption{The considered system model.}
\label{fig: System_model}
\end{figure}

As shown in Fig. \ref{fig: System_model}, the considered system model consists of the following entities:
\begin{itemize}
    \item The charging coordinator (CC): The CC receives charging requests from EVs, computes charging schedules and sends them to the EVs. It has to make sure that the scheduled charging amounts do not exceed the available charging energy. 
    \item Electric vehicle $(EV)$: The EVs send charging requests containing information, such as battery SoC, to the CC that uses them to compute the charging schedules of the EVs. 
   \item Collector: It is responsible for collecting the EVs' charging requests at a given area and sending them to the CC. It also receives charging schedules from the CC and forwards them to the EVs in its area.
    \item Charging point (CP): It provides the needed electrical power charging to the EV.
\end{itemize}

As shown in Fig. \ref{fig: System_model}, the charging coordination mechanism is executed by the CC for each geographic area that is usually connected to one electrical bus. The total amount of power charging of the EVs in each area should not exceed the maximum charging capacity known to the CC. The idea is that time is divided into slots, e.g., 30 minutes for each slot, and the charging coordination mechanism is run at the beginning of each time slot. When an EV needs to charge, it should send a charging request to CC including some information such as SoC and time to complete charge (TCC) \cite{electric_vehicles_scheduling}. TCC is the expiry time of the charging request. After receiving the charging requests of EVs in a certain area by the CC, if the total charging demand does not exceed the charging energy capacity, all EVs can charge. On the other hand, in case that the total charging demand exceeds the charging capacity, the CC should run a charging coordination mechanism to select a subset of the EVs to charge in this time slot and defer the charging of the other EVs to the next time slots. The mechanism should prioritize the requests and select the high-priority requests for charging. Typically, more priority is given to the EVs that have low SoC. In the literature, there are different ways to select the subset of EVs that should charge \cite{RealTimeCharging, 7031457, 7564967}.

For the threat model, we focus on attacks launched by lying EVs that send false data in their charging requests to illegally gain higher priority in charging. 
In this paper, we propose different ways to report false data by EVs. We train a machine learning detector that takes the SoC values reported by an EV and classifies the EV to honest or liar.

\section{Evaluation of the Impact of Lying EVs}
\label{Problem Formulation}

In this section, we run multiple experiments to evaluate the impact of reporting false SoC on the charging coordination mechanism. Specifically, we are interested in investigating how this attack: (1) benefits lying EVs, (2) harms the honest EVs, and (3) degrades the performance of the charging coordination mechanism.

\subsection{Setup}
We use MATLAB~\cite{matlab} to simulate the following charging coordination scenario. Define a set of $n$ EVs ($\mathbb{E}=\{EV_1,EV_2,\ldots,EV_n\}$), a set of days ($\mathbb{D}=\{1,\ldots,d\}$), and a set of time slots of equal time periods ($\mathbb{T}=\{1,\ldots,T\}$), for an $EV_i \in \: \mathbb{E}$, the SoC value at day $d \in \mathbb{D}$ and time $t \in \mathbb{T}$ is donated by $S_i(d,t)$. 
At time $t$, $EV_i$ sends a charging request that has some information such as the SoC of the EV $S_{i}(d,t)$ to the CC. A charging coordination mechanism that uses the EVs' charging requests is executed to decide the EVs that can charge in this time slot. However, lying EVs report false SoC value (i.e., lower value), denoted by $\beta*S_i(d,t)$ where $0\leq\beta<1$, to the CC for the purpose of gaining higher charging priority illegally to charge quickly without any deferral.
 
 In our experiments, we assume that the maximum capacity of each EV's battery is 200 units, the total number of EVs is 100, and the simulation is run for a total of 30 time slots. In addition, the simulation is run using two charging capacities per time slot, which are $1080$  and $2160$ units of power, to evaluate the impact of reporting false SoC on the charging coordination mechanism at two different levels of available charging supply. Initially, the SoC of all vehicles is assumed to be $0.5$ (i.e., the battery of each EV is half charged) with a time to complete charge (TCC) equal to four. In our experiments, we run the charging coordination mechanism proposed in~\cite{7564967} that is based on knapsack algorithm, as illustrated in Algorithm~\ref{KnapSack}. The priority index ($PI$) of each charging request is a number between $0$ and $1$, where the priority of the request increases as $PI$ increases. $PI$ is calculated as follows:
\begin{equation*}
PI =\varepsilon f_1(SoC) + (1 - \varepsilon)  f_2(TCC),
\label{eq:30}
\end{equation*}
where $\varepsilon$ is a factor that is between $0$ and $1$, and is used to determine the weights given to SoC and TCC. $f_1()$ and $f_2()$ are two functions that map SoC and TCC, respectively, to numbers between $0$ and $1$.  
We assume that $\varepsilon$ is $0.5$ and $f_{2}(x)$ is $0.4$ when $0<x\leq4$. For simplicity, we assume that $f_{1}(x)$ is a binary function that gives $1$ when $0.4<x\leq1$ and $0.1$ otherwise. In addition, the SoC reported by an honest $EV_{i}$ is $S_{i}(d,t)$ while the SoC reported by a lying $EV_{i}$ is $\beta*S_{i}(d,t)$ where $0\leq\beta<1$ and $S_{i}(d,t)$ is the true SoC. In the experiments, we vary $\beta$ to study different cheating behaviors. We assume that EVs need to fully charge their batteries, i.e., until SoC is 1.

\SetAlFnt{\small}
\begin{algorithm}[!h]
\SetKwProg{fn}{function}{}{}
\SetKwData{NumOfUpdatedObjects}{numOfUpdatedObjects}
\SetKwIF{If}{ElseIf}{Else}{if}{}{else if}{else}{end if}
\SetKwIF{For}{ElseIf}{Else}{for}{}{else if}{else}{end For}
\SetKwFunction{KnapSack}{KnapSack}

\fn{Charging Coordination ($SOC$, $TCC$)}{ 

$C_{sys}$ $\leftarrow EnergyCapacity$
\BlankLine
$\varepsilon \leftarrow 0.5$
\BlankLine
\tcp{$X$: the amount of energy each EV can charge in this time slot}
\For{$i \in X$}{
$X_{i} \leftarrow 0$
}
\BlankLine
\tcp{$P$: the charging amount requested by EVs}
\For{$i \in SOC$}{
$P_{i} \leftarrow 1-SOC_{i}$
}
\BlankLine
\tcp{$PI$: priorities of EVs}
\For{$i \in SOC$}{
\begin{equation*}
PI_{i} \leftarrow \varepsilon f_1(SOC_{i}) + (1 - \varepsilon)  f_2(TCC_{i})
\end{equation*}
}
\BlankLine
$A \leftarrow PI_{1}/P_{1} \geq PI_{2}/P_{2} \cdots \geq PI_{n}/P_{n}$
\BlankLine
\For {$i \in A$}{
\If {$P_{i} \leq C_{sys}$}{
\BlankLine
$X_{i} \leftarrow P_{i}$
\BlankLine
$C_{sys} \leftarrow C_{sys} - P_{i}$
\BlankLine
$A \leftarrow A - i$
\BlankLine
}
}
\tcp{if there is remaining power, provide it to the EV with the highest priority}
$L \leftarrow argmax_{A}PI$
\BlankLine
$X_{L} \leftarrow C_{sys}$
\BlankLine
return X
}




\caption{Charging coordination mechanism used in \cite{7564967}.}
\label{KnapSack}
\end{algorithm}




\subsection{Metrics} 
In this subsection, we define the key metrics that are used for evaluating the impact of reporting false SoC by EVs to the charging coordination mechanism. 
The first metric is the \textit{probability that lying EVs get charged} before the expiry of their charging requests. The second metric is the \textit{probability that honest EVs can charge} in the existence of lying EVs. The third metric we will measure is the \textit{amount of unused energy}. As discussed earlier, the lying EVs report lower SoC to gain higher priority and this may result in assigning more power to the EVs than what they need which results in unusing some charging energy, i.e., the total amount of energy charged by EVs is below the available charging capacity. Note that, the EVs pay for charging based on the actual consumption (not the assigned charging power by CC) measured by meters connected to the charging outlet. Therefore, it is possible that some power assigned by CC to the lying EVs are not used and thus not assigned to other EVs which definitely degrades the performance of charging coordination mechanism because it increases the likelihood that EVs' charging requests are expired without getting charged.
 
 \subsection{Simulation Results}

\begin{figure*}[t!]
\centering
\subfigure[$\beta=0.2$ and total charging energy is $1080$.]{
\includegraphics[width=3.1in]{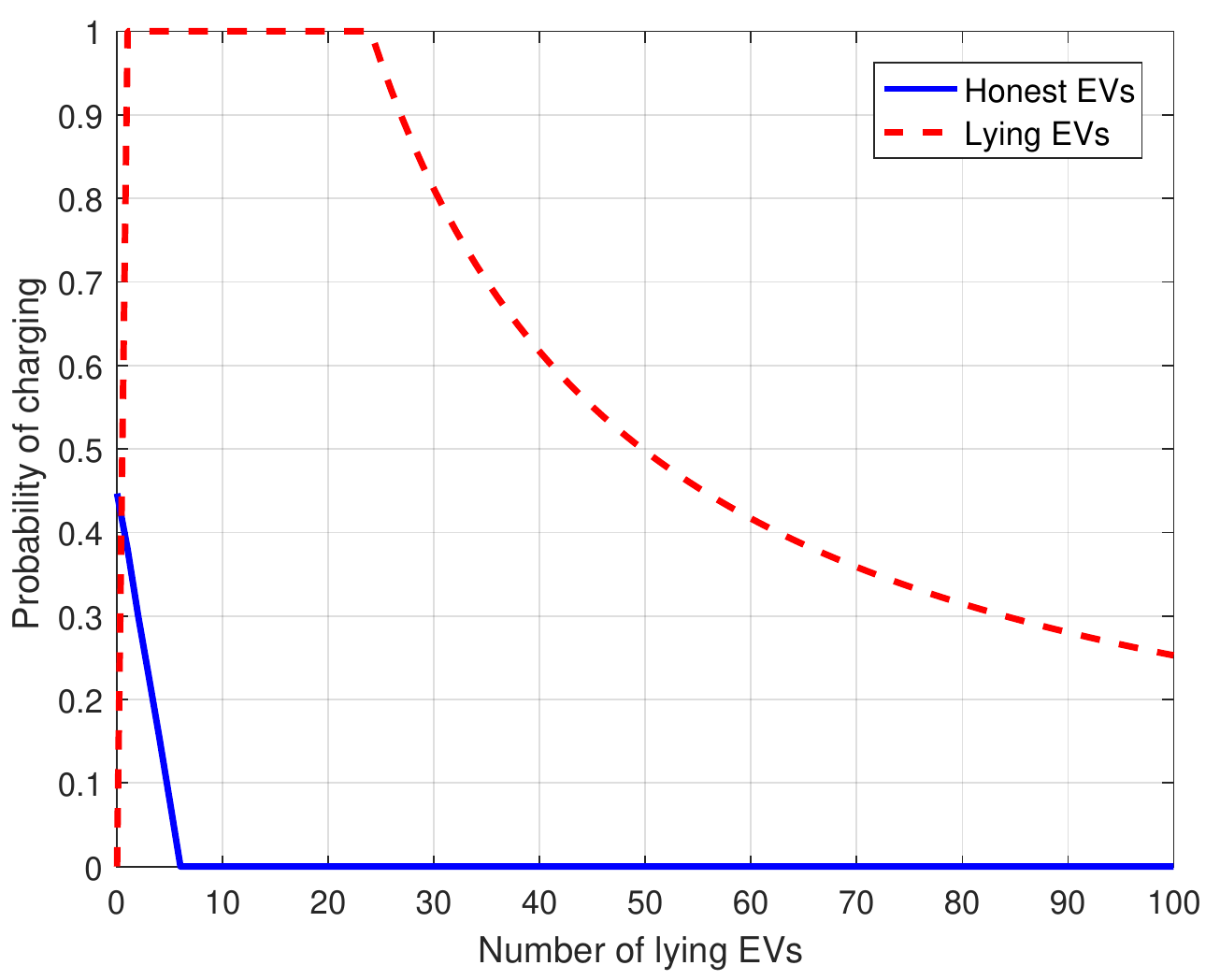}
\label{subfig:1}
}
\subfigure[$\beta=0.8$ and total charging energy is $1080$.]{
\includegraphics[width=3.1in]{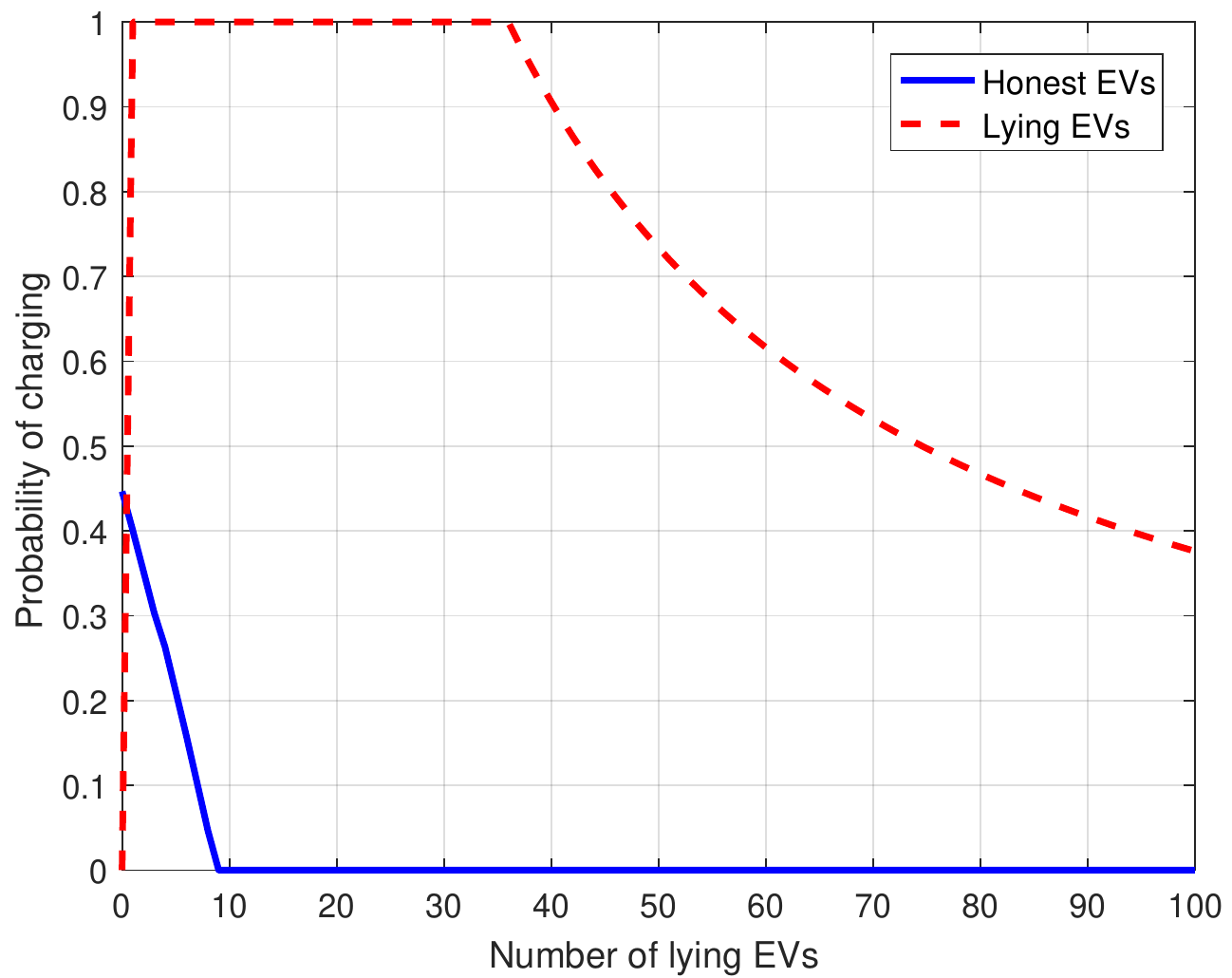}
\label{subfig:2}
}\\
\subfigure[$\beta=0.2$ and total charging energy is $2160$.]{
\includegraphics[width=3.1in]{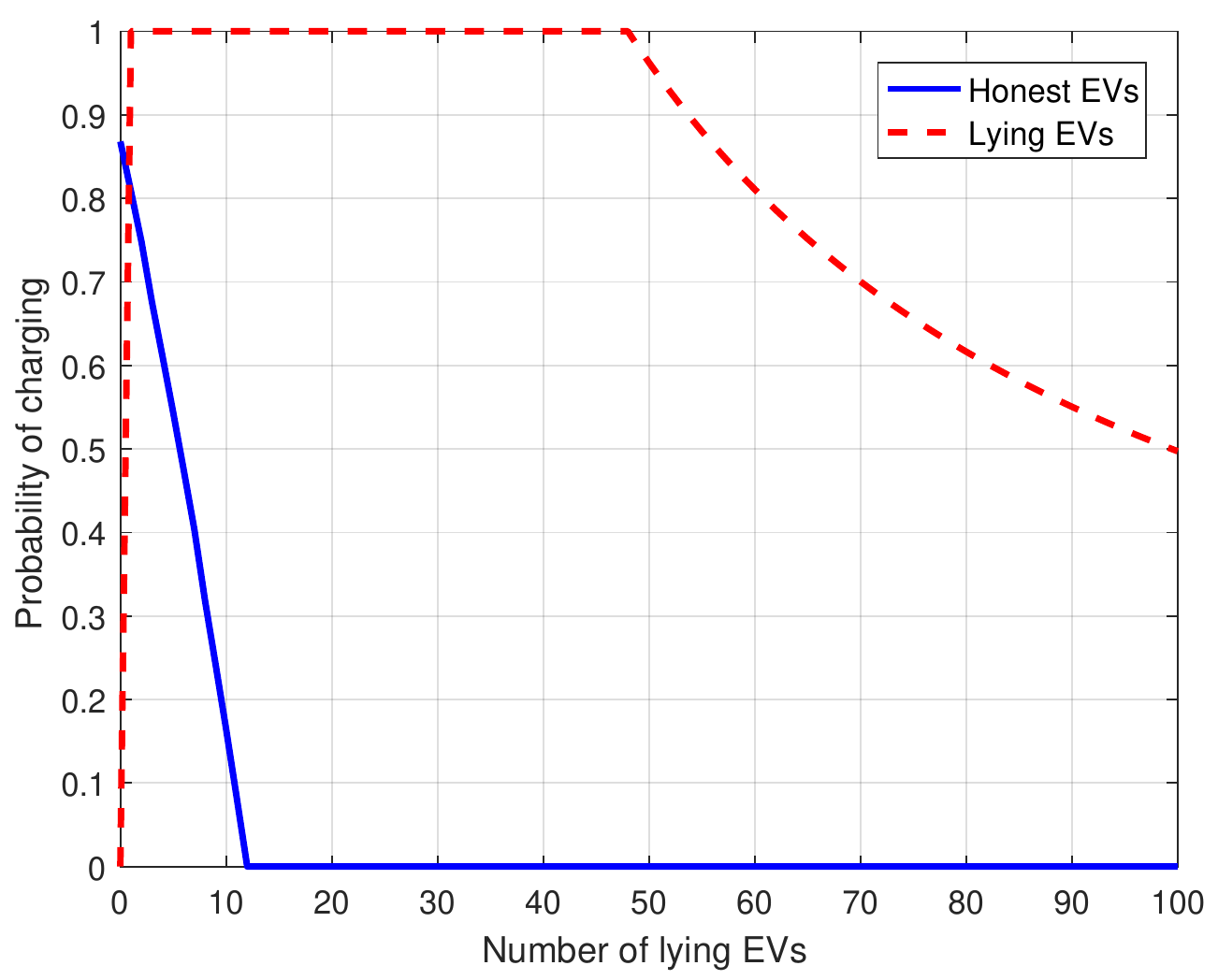}
\label{subfig:3}
}
\subfigure[$\beta=0.8$ and total charging energy is $2160$.]{
\includegraphics[width=3.1in]{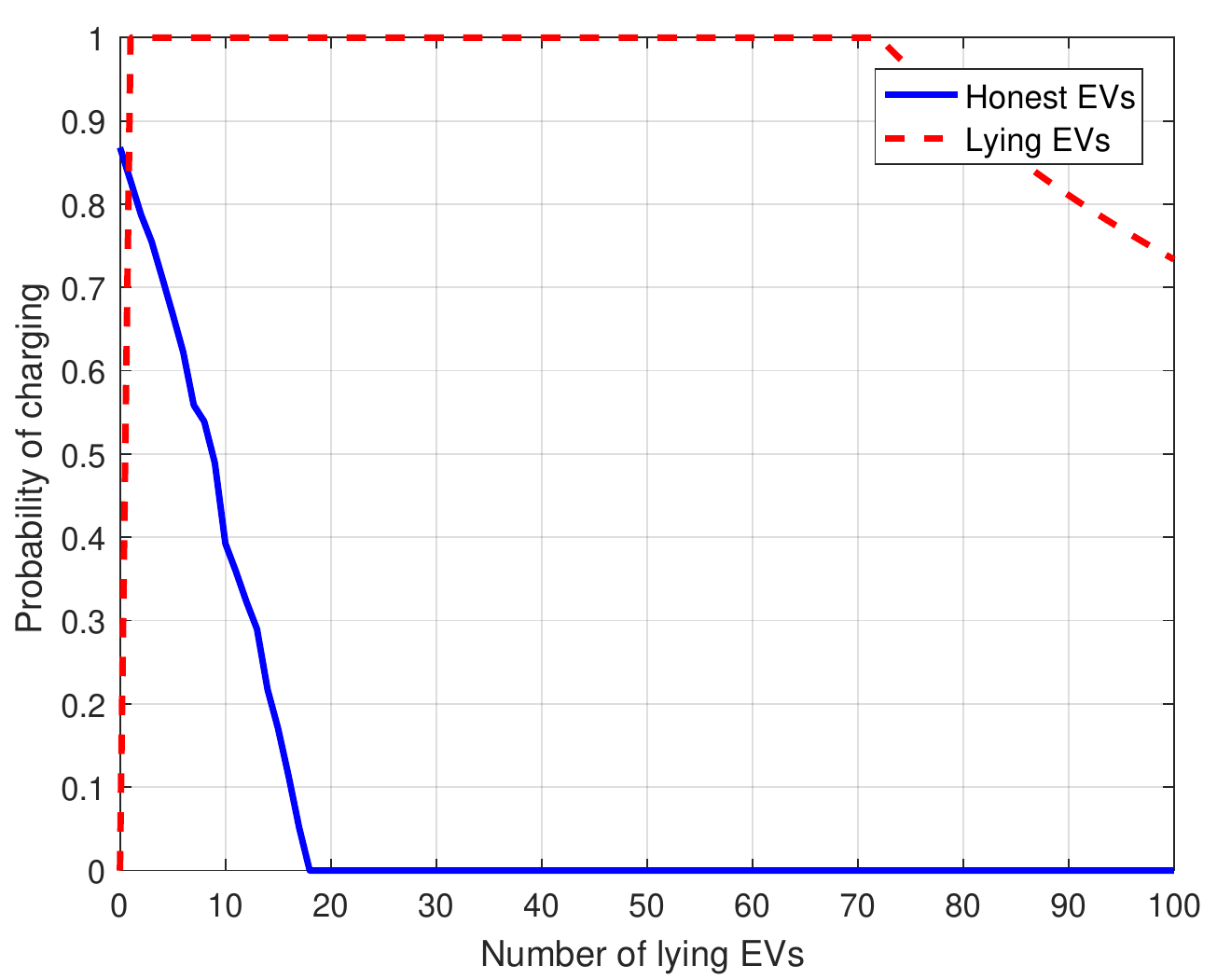}
\label{subfig:4}
}
\caption{The probability of selecting honest and lying EVs for charging at different values of $\beta$ and total charging energy.}
\label{probability}
\end{figure*}

Fig.~\ref{probability} gives the probability of selecting honest and lying EVs for charging at different values of $\beta$ and total charging energy. The probability of selecting lying EVs for charging is the number of lying EVs that fully charge before their charging requests expire divided by the total number of lying EVs. Similarly, the probability of selecting honest EVs for charging is the total number of honest EVs that are selected for charging before their charging requests expire divided by the total number of honest EVs. The figure shows that as the number of lying EVs increases, as the probability of selecting honest EVs for charging decreases. This is because more lying EVs are selected for charging because they have higher (false) priority. At a certain number of lying EVs, the probability of selecting honest EVs becomes zero. This is because at this point, none of the honest EVs are selected and the lying EVs are always selected. For instance, when $\beta$ is $0.2$ and the total charging energy is $2160$, $12$ EVs are selected for charging in each time slot. So, when the number of lying EVs is at least 12, none of the honest EVs are selected and the lying EVs are always selected because they have higher priority. 

It can also be seen that the probability of selecting lying EVs is consistently one until the lying EVs reaches to a certain number, and after that, the probability starts to decrease. The justification for this behavior is that when the system has many lying EVs, some lying EVs may not be selected for charging before their charging requests expire. It can also be seen that as $\beta$ decreases, fewer number of honest EVs are selected for charging. For instance, from Figs.~\ref{subfig:3} and~\ref{subfig:4} where $\beta$ is $0.2$ and $0.8$, when the numbers of lying EVs are 12 and 18, respectively, the number of honest EVs selected for charging is zero. In addition, fewer number of EVs are selected at higher $\beta$ because as $\beta$ decreases, more power is assigned to lying EVs so less power is left to honest EVs, which results in selecting fewer number of EVs without exceeding the total charging energy. Morever, by comparing Figs.~\ref{subfig:1} and ~\ref{subfig:3} and Figs.~\ref{subfig:2} and ~\ref{subfig:4}, it can be seen that by reducing the total charging energy from $2160$ to $1080$, fewer number of EVs are selected for charging in each time slot. This results in selecting fewer number of honest EVs, and also reducing the probability of charging lying EVs at a smaller number of EVs. Therefore, lying EVs have a more severe impact as the available charging energy reduces. 

\begin{figure}[!t]
\centering
\captionsetup{justification=centering}
\includegraphics[width=3.1in]{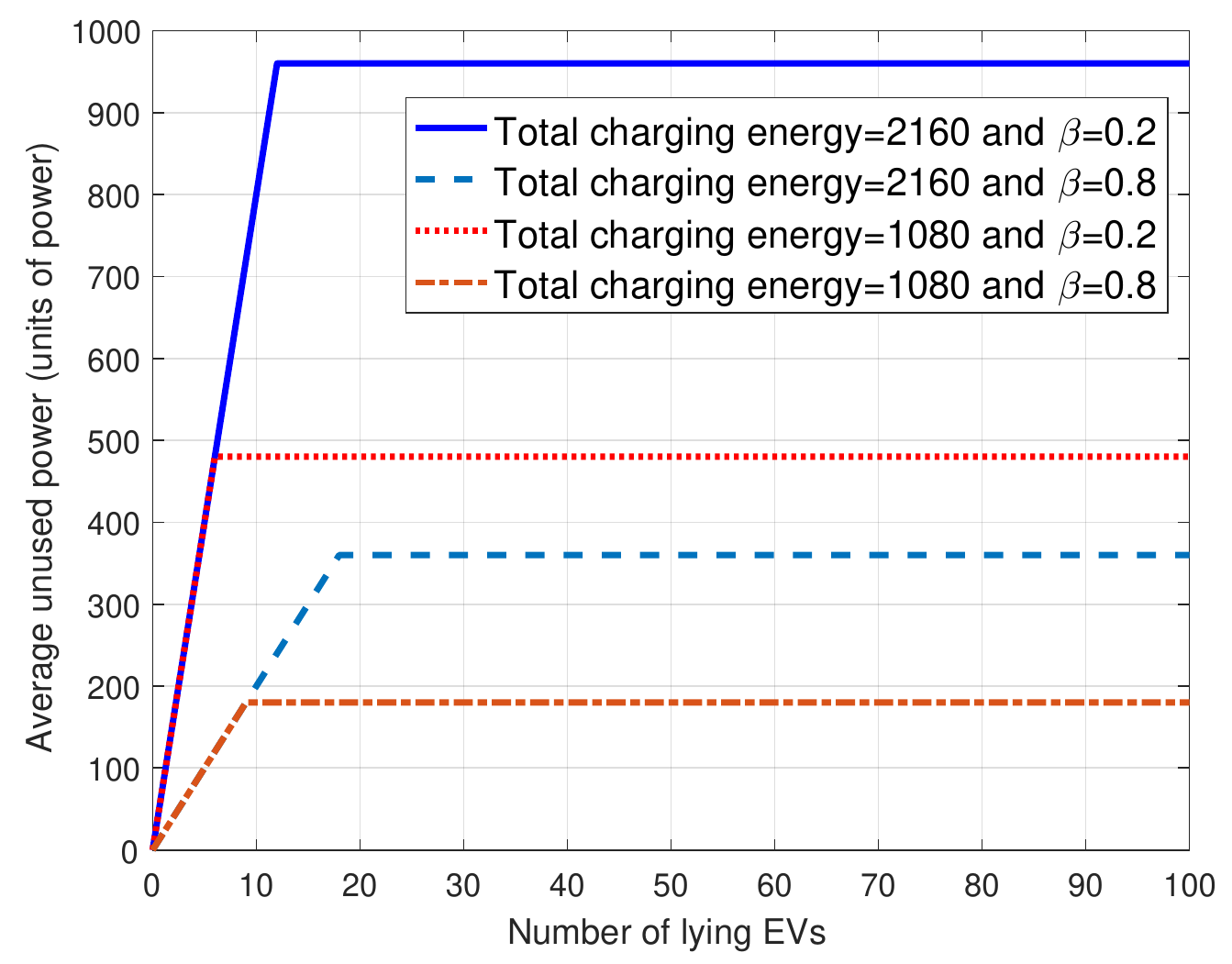}
\caption{The average unused power per time slot.}
\label{fig: APG}
\vspace{-4mm}
\end{figure}

Fig.~\ref{fig: APG} gives the average amount of unused power versus the number of lying EVs. As shown in the figure, as the number of lying EVs increases, more power is unused because more lying EVs are selected for charging, and as explained earlier, charging energy is unused because more power is assigned to lying EVs than what they actually need. At a certain number of lying EVs, the unused power is saturated. This is because all the selected EVs for charging at this number are lying ones. It can also be seen that as $\beta$ decreases, i.e., reporting more lower SoC, more power is unused because more power is assigned to the lying EVs. Finally, comparing the two cases of total energy, it can be seen that more power is unused when the total charging energy increases from $1080$ to $2160$ because more lying EVs are selected for charging.

\section{Preliminaries}
\label{Preliminaries}
In this section, we present a brief description of deep learning and optimization techniques that will be used in the design of our detectors.
\subsection{Deep Learning}
Deep learning is a special case of neural networks characterized by more than one middle hidden layer. The traditional structure of neural networks generally consists of three basic layers: input, output and one or more hidden layers \cite{8233155}.
Deep networks can extract complex attributes (features) from unstructured data which is a complex task especially when the attributes are hard to obtain \cite{Shokri2015}. Deep learning could be categorized into \textit{supervised} and \textit{unsupervised}. In \textit{supervised deep learning}, a labeled dataset is used to compute a deep learning model. Examples of such deep learning networks are the multi-layer perceptron (MLP) \cite{ijcai2019-647}, convolutional neural network (CNN) \cite{NIPS2017_6975}, and recurrent neural network (RNN) \cite{ha2018worldmodels}. Regrading the \textit{unsupervised deep learning}, its main objective is to cluster unlabeled data with similar features together, Deep autoencoder network \cite{8264962} is an example of these kind of networks.

Backpropagation is the algorithm that is used to update the network weights according to the error between the real label $y$ of an input sample and the output of the deep network $z$ \cite{BackPropogation} using Gradient descent (GD) \cite{Shokri2015} technique. Batch gradient descent is a modification on the traditional GD, in which the weights are computed by averaging over the whole dataset. In stochastic gradient descent (SGD) \cite{45428}, the weight gradients are averaged over a small subset of the whole training dataset that is called the mini-batch. From the numerous categories of the deep learning models that have been stated before, our focus in this paper is on the supervised category and specifically on the deep recurrent neural network (RNN) and the multi-layer perceptron (MLP) that will be described in more details in the following subsections. 

\begin{figure}[!t]
\centering
\captionsetup{justification=centering}
\includegraphics[width=1\columnwidth]{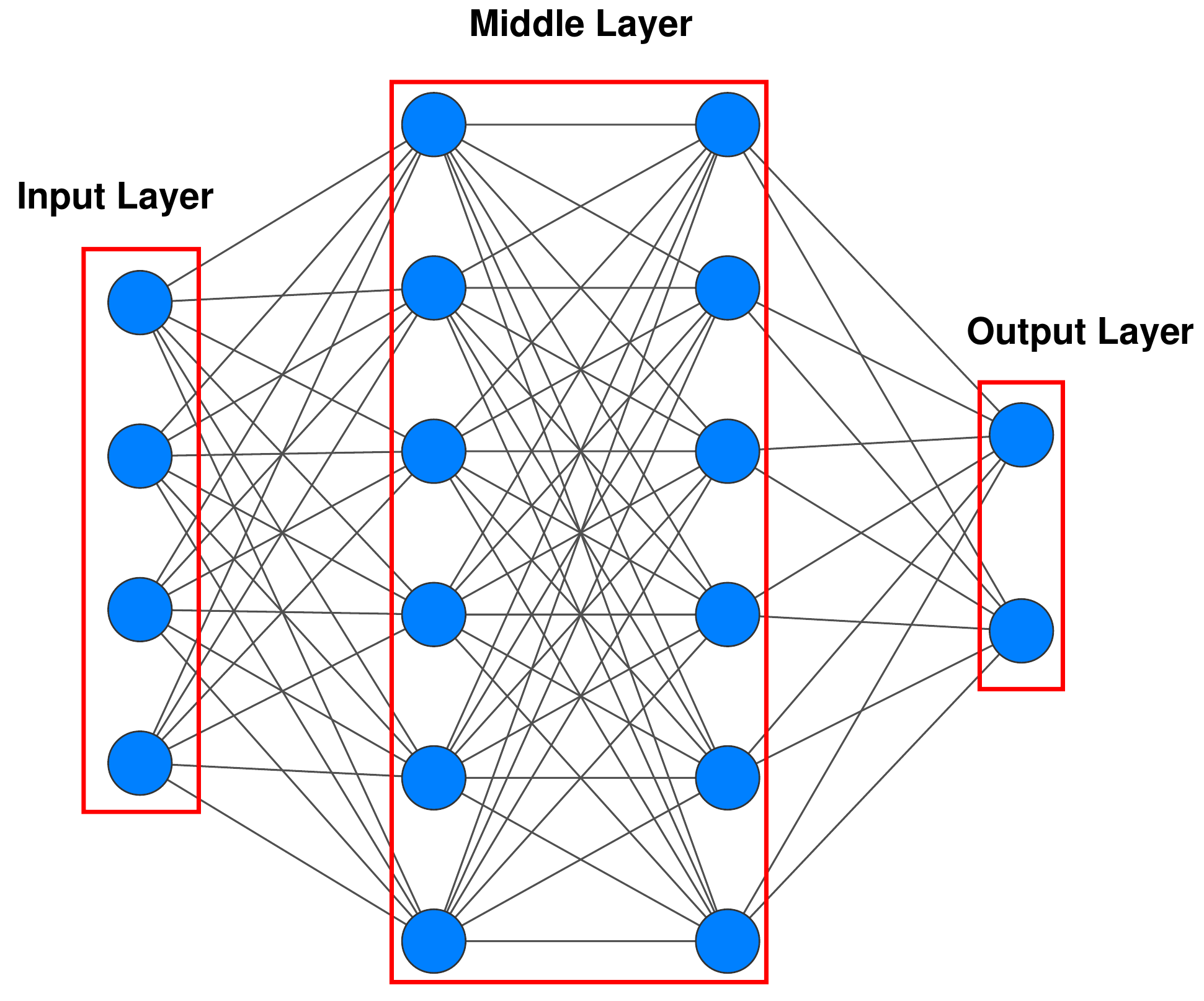}
\caption{Architecture of multi-layer perceptron (MLP) network.}
\label{fig: MLP}
\vspace{-2mm}
\end{figure}

\subsection{Multi-Layer Perceptron (MLP)} MLP is a type of deep neural networks (DNN) \cite{ijcai2019-647}, 
which includes an input layer for receiving the input, an output layer for producing the decision and one or more of the middle layers as shown in Fig. \ref{fig: MLP}. The middle layers are the ones that are responsible for the non-linear operations that are done on the input to compute an output decision on the last layer. MLP is usually used to solve non-linear classification problems or problems that have a non-linear separable set of patterns. This non-linearity is achieved by using a given function, called activation function, in each of the neurons of the hidden layer and at the output layer. 
The most commonly used activation functions are Sigmoid, Tanh, Threshold and Softmax.

\begin{figure}[!t]
\centering
\captionsetup{justification=centering}
\includegraphics[width=0.8\columnwidth]{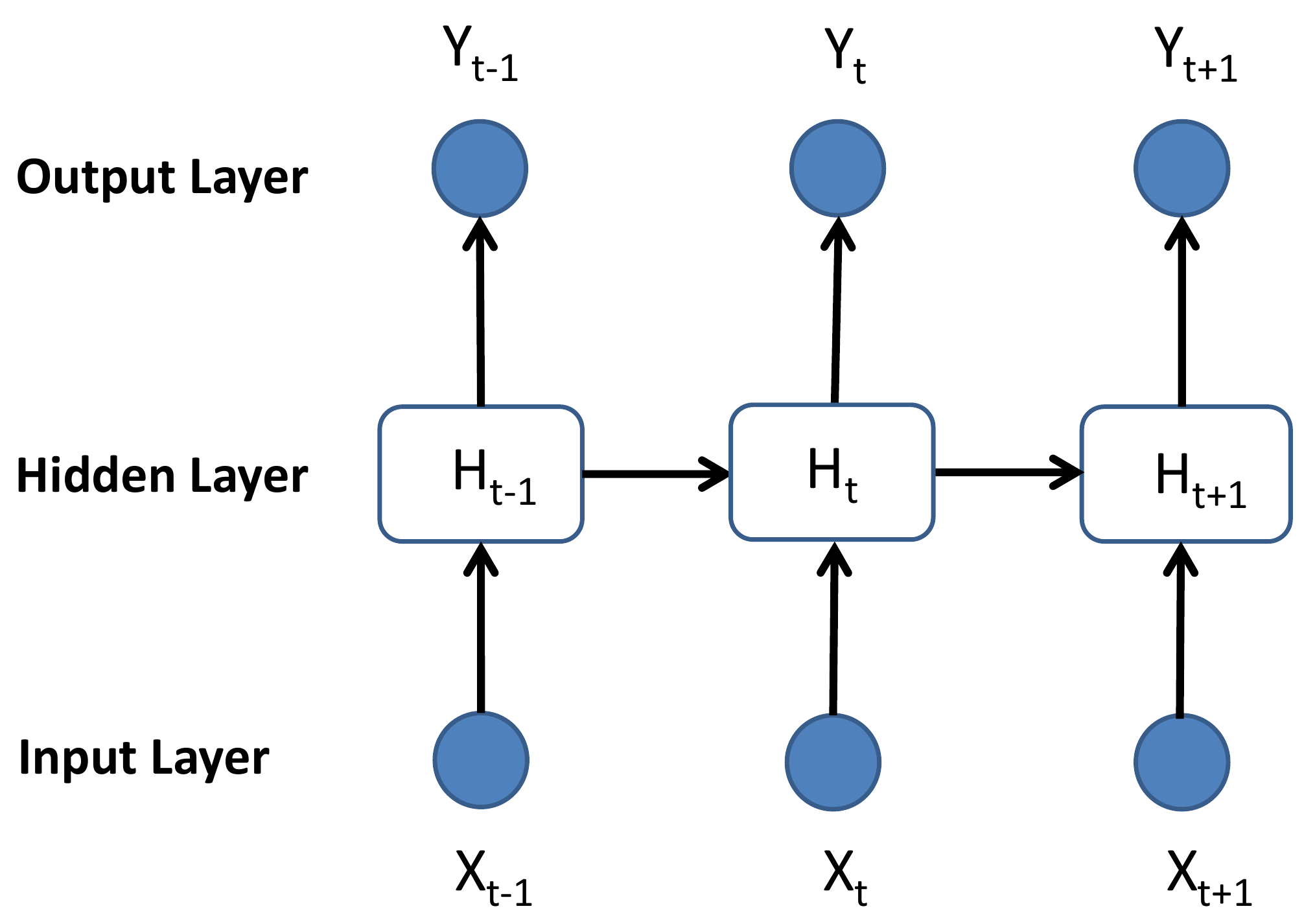}
\caption{Architecture of deep recurrent neural network (RNN).}
\label{fig: RNN}
\vspace{-4mm}
\end{figure}

\subsection{Recurrent Neural Network (RNN)} 
Both inputs and outputs are independent in conventional neural networks, but in applications that need to predict the next word of a sentence, the preceding words are required. Thus, as shown in Fig. \ref{fig: RNN}, an RNN \cite{7846305} uses hidden layer(s) to solve this problem.  RNN's main and most significant function is the hidden state ($H_{t}$), which recalls some information about the sequence of the data \cite{pmlr-v37-gregor15}. 
The node-to-node connections in RNNs form a directed graph along a temporal sequence and this reflects the temporal dynamic behavior.  Unlike feed-forward neural networks, input sequences can be processed by RNNs using their hidden state (memory) that makes them perfect for tasks such as unsegmented handwriting recognition or recognition of voice \cite{7846306}. Feedback loops are used to process the input data sequence producing the final output, that can also be a data sequence. Such feedback loops allow the retention of information and this effect is often represented as memory that preserves all the measured parameters and links the inputs together and allows sequential and temporal information to be processed by RNNs. The algorithm that is used by the RNNs  to loop information back into the network throughout the computational process for updating the weights through the network is backpropagation through time (BPTT) \cite{NIPS2016_6039}. 

Long sequences RNNs have a difficulty to carry (remember) information from earlier time steps to the later steps. Thus, during the parsing process of a long text paragraph for prediction purposes, RNNs may ignore important information from earlier steps. In addition, during the backpropagation technique, RNNs could suffer from the vanishing gradient problem, that is when the gradient weight becomes too small as it back propagates through time so that it does not contribute enough during the learning process. Accordingly, the learning process of the layers that get small gradient update, are stopped and usually those layers are the earlier layers. As these layers are not involved in the learning process, RNNs may forget important details from earlier stages, as mentioned before, when it encounters long sequences, and thus having what we call, short-term memory.
Accordingly, two special kinds of the RNN are introduced in literature to solve this problem, namely, \textit{Long Short Term Memory (LSTM)} \cite{lstm}, and \textit{Gated Recurrent Unit (GRU)} \cite{8053243}. 
Their internal structure contains internal mechanisms, called gates, which can regulate the information flow. The learning process of these gates involves teaching what is important to keep and what can be thrown away, so that during the parsing of long sequences, relevant information can be passed down the long chain for making predictions. These two RNNs are commonly used in different applications such as: text generation, speech recognition, caption generation of videos and speech synthesis. The GRU is a variation of the LSTM with a reduced set of gates while having the same ability to address the vanishing gradient problem.

\subsection{Genetic Optimization Technique}
Genetic algorithm (GA) \cite{MAULIK20001455} is a searching strategy that is used to find the best or fittest solution for a given (optimization) problem. GA represents a population of candidate solutions using a genetic representation. After that, GA moves toward a better solution using evolutionary concepts such as the natural selection and survival of the fittest.  

Five phases are considered in GA which are: initial population, fitness function, selection, crossover, and mutation. The process begins with a set of individuals which are considered as a population, where each individual is a solution to the problem we want to solve. In addition, an individual (solution) is represented in a chromosome form which is a string of genes where each gene represents a variable (parameter) that characterizes this individual.  Each individual in the population is evaluated using the fitness function to give the individual a fitness score. The individuals with the highest fitness are selected in the selection phase and their genes are passed to the next generation. The crossover phase is done on each pair of selected individuals where a random crossover point is chosen from the genes and an offspring solution is created by exchanging the genes of the two parents among themselves until the crossover point is reached. Consequently, a new offspring solution is added to the population. 

From the newly created offspring individuals, some could be subject to the mutation process with a low random probability. This implies that some of the genes of the chromosomes can be flipped.  Mutation occurs to maintain diversity within the population and prevent premature convergence. In addition, a small portion of the fittest solutions are copied into the next generation. This process is called elitism, and could have a dramatic impact on the performance by ensuring that the GA does not waste time in re-discovering previously discarded partial solutions. Elitism guarantees that the solution quality obtained by the GA does not degrade from one generation to the next. The whole process is repeated for every population until the GA converges.  At this point the GA provides a set of solutions to our problem. Convergence means that the algorithm does not produce an offspring that is notably different from the previous generation.

\section{EV's Dataset}
\label{Dataset}
In this section, we explain how we created a dataset for charging coordination application using real driving traces of vehicles and technical information released by an EV manufacturer. Although, our focus is to create a dataset that is used to train a detector to identify lying EVs, our dataset can also be used for other applications such as prediction of energy demand. 

{\renewcommand{\arraystretch}{1.8}
\begin{table}[t]
\centering
\caption{Operational data of Kia Soul EV.}
\label{tab:Operational-Data-Of}
\begin{tabular}{|
 >{\centering\arraybackslash}m{4.5cm}| >{\centering\arraybackslash}m{2.5cm}|}
\hline 
Brand & Kia \\
\hline 
Model & Soul\\
\hline 
Battery Capacity (KWh) & 64\\
\hline 
Ave. Electric Range (Mi) & 230 \\
\hline 
Max Charge Rate (KW) & 7.2\\
\hline 
Power Consumption (Wh/mi) & 275 \\
\hline 
\end{tabular}
\end{table}
}


\subsection{Honest Dataset Creation}
For creating the honest dataset, we use driving routes of 536 taxis in San Francisco, CA, released in \cite{TaxiCabs}.
The data recording began on May 17, 2008 and finished on June, 10, 2008. For each taxi, each data row contains a time-stamp, latitude, and longitude. 
In our dataset, we consider Kia Soul EV\cite{kiaSoul}.
In addition to using the real routes in \cite{TaxiCabs}, we also use information released by Kia company. These information are summarized in Table \ref{tab:Operational-Data-Of}. 
As shown in the table, the EV uses a charging technique \cite{kiaSoul}, in which the grid provides the EV with the required amount of electrical power using the alternating current (AC) which is converted into a direct current (DC) using the EV's own rectifier (converter). 

In our dataset, the SoC is linearly dependant on the distance driven by the EVs \cite{DBLP:journals/corr/abs-1802-04931}. To compute the SoC at every minute for each EV, we initialize the SoC value randomly, and then for every minute, we check if the EV is moving or not. If it is moving, then we first compute the distance using the EV speed and the time, and then we compute the power consumption of the EV by multiplying this distance by the power consumption rate given in Table \ref{tab:Operational-Data-Of}, and finally update the value of the SoC by subtracting the ratio of the power consumed and the maximum battery capacity from the current SoC value. If the EV is not moving then we check if it is charging or not. If it is charging, then the SoC value is updated by adding the ratio of the amount of the power drawn and the EV's capacity. If the EV is not charging then the new SoC value is the same as the old value.

A benign dataset is created for each EV of the 536 EVs that describes its normal behavior. The behavior of each EV is introduced using 24 rows in the dataset representing the normal behavior of the EV in 24 days. Each row contains $T$ features, resembles the SoC values reported every $\tau$ minutes for that EV. In this paper, $\tau$ is selected to be 30 minutes producing a set of ($T=48$) features in each row. A total of $12,864$ data samples are created to represent the benign dataset. 

\begin{figure}[!t]
\centering
\captionsetup{justification=centering}
\includegraphics[width=\columnwidth]{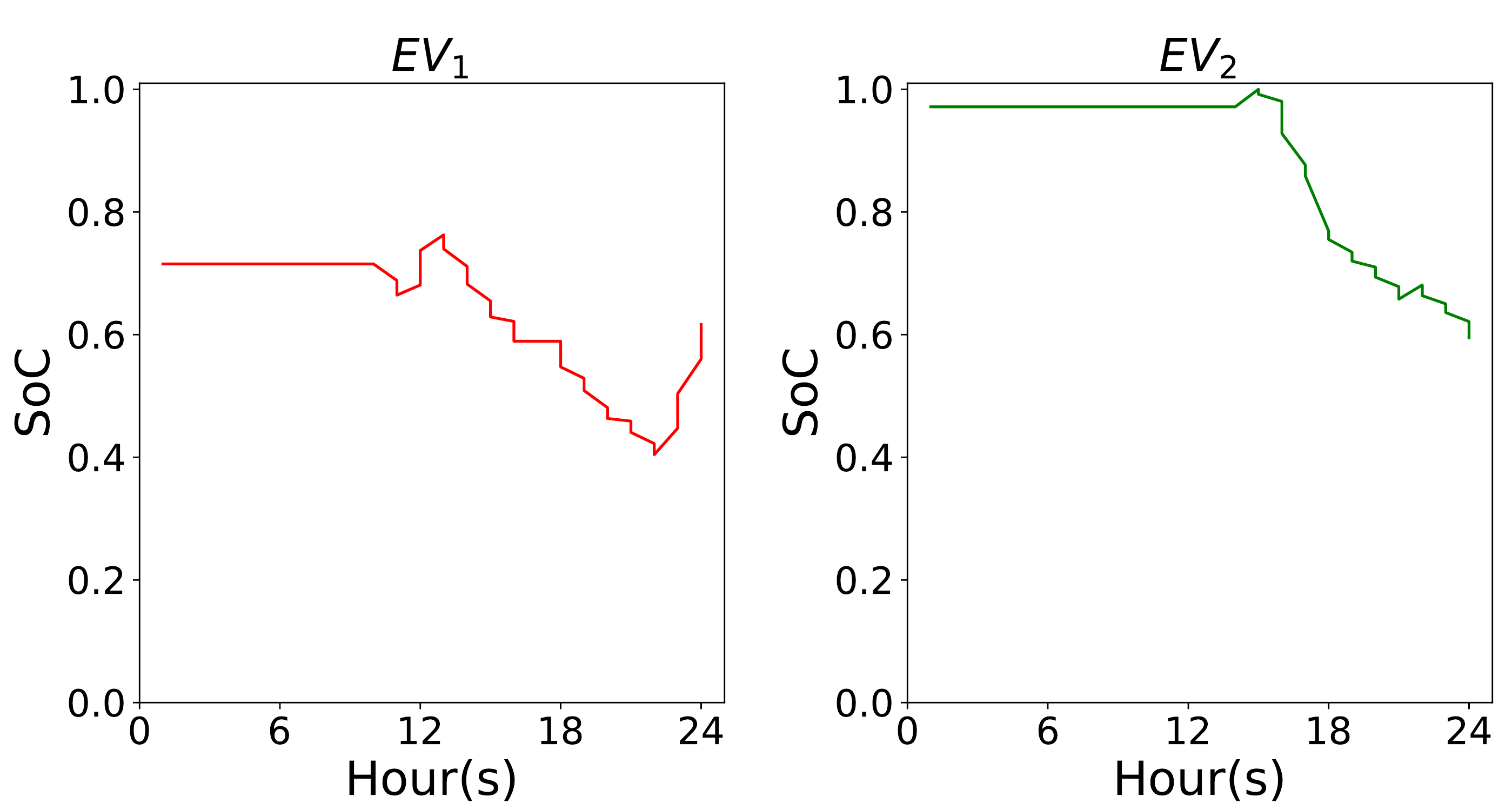}
\caption{The SoC readings of two EVs for one day from the dataset.}
\label{fig: one_day_four_users}
\end{figure}

Fig. \ref{fig: one_day_four_users} shows the SoC values of two EVs for one day. As observed from the figures, the SoC initial values are different for the two EVs, and the SoC values start to decrease at a certain hour which indicates that the EVs are moving, however, at some time the SoC values start to increase indicating that the EVs are charging at these points of time. 

\begin{figure}[!t]
\centering
\captionsetup{justification=centering}
\includegraphics[width=\columnwidth]{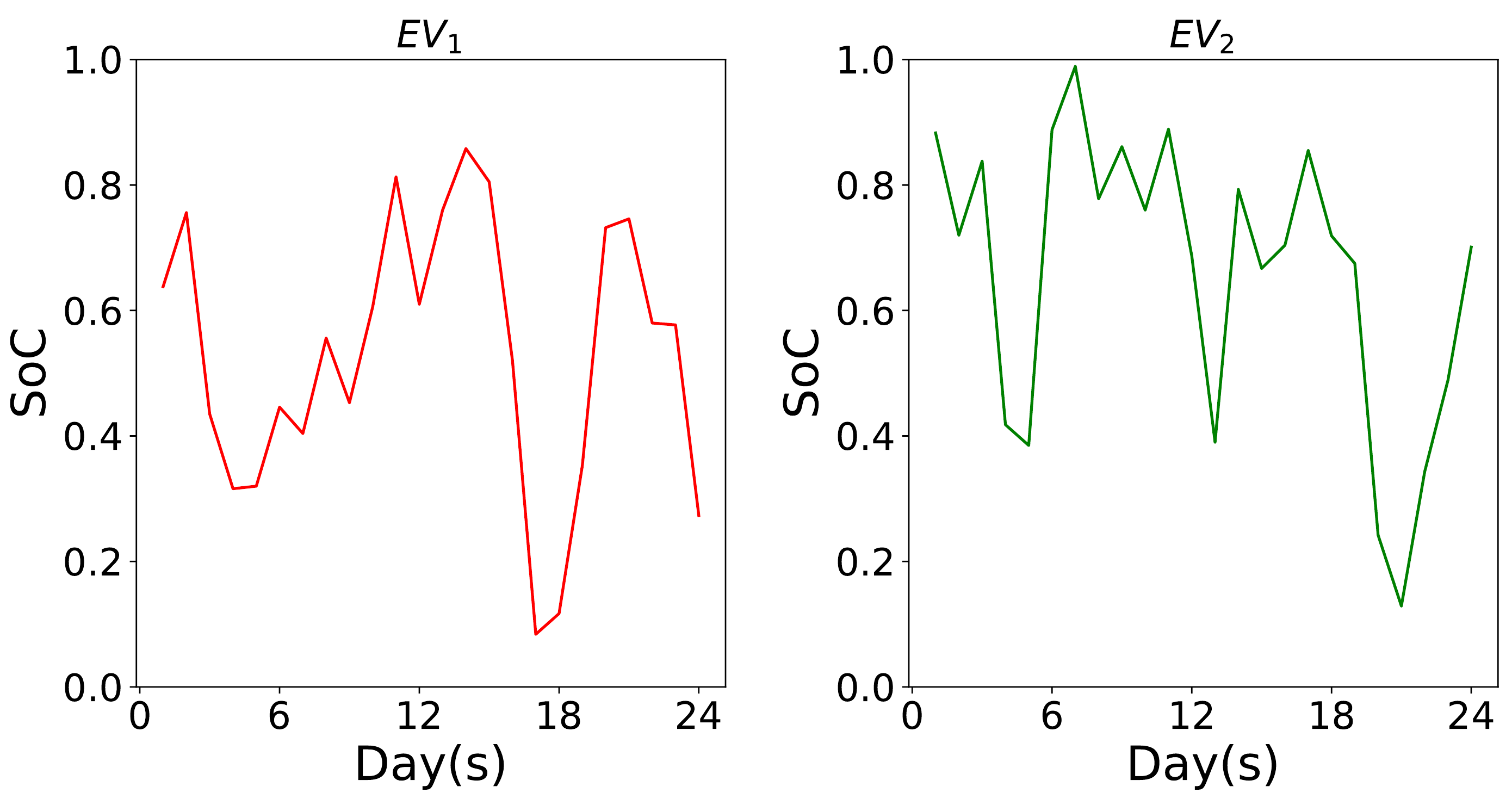}
\caption{The average SoC values of two EVs for several days from the dataset.}
\label{fig: one_week_four_users}
\end{figure}

Fig. \ref{fig: one_week_four_users} shows the SoC values of two EVs through 24 days. As shown in the figures, for each of the two EVs, there is a different SoC pattern indicating charging or not charging, and moving or still. 
This indicates that each EV has a normal behavior consistent with its pattern of SoC, and our detector will try to detect the EVs that deviates from their normal behavior and label them liars.

\begin{figure}[!t]
\centering
\captionsetup{justification=centering}
\includegraphics[width=\columnwidth]{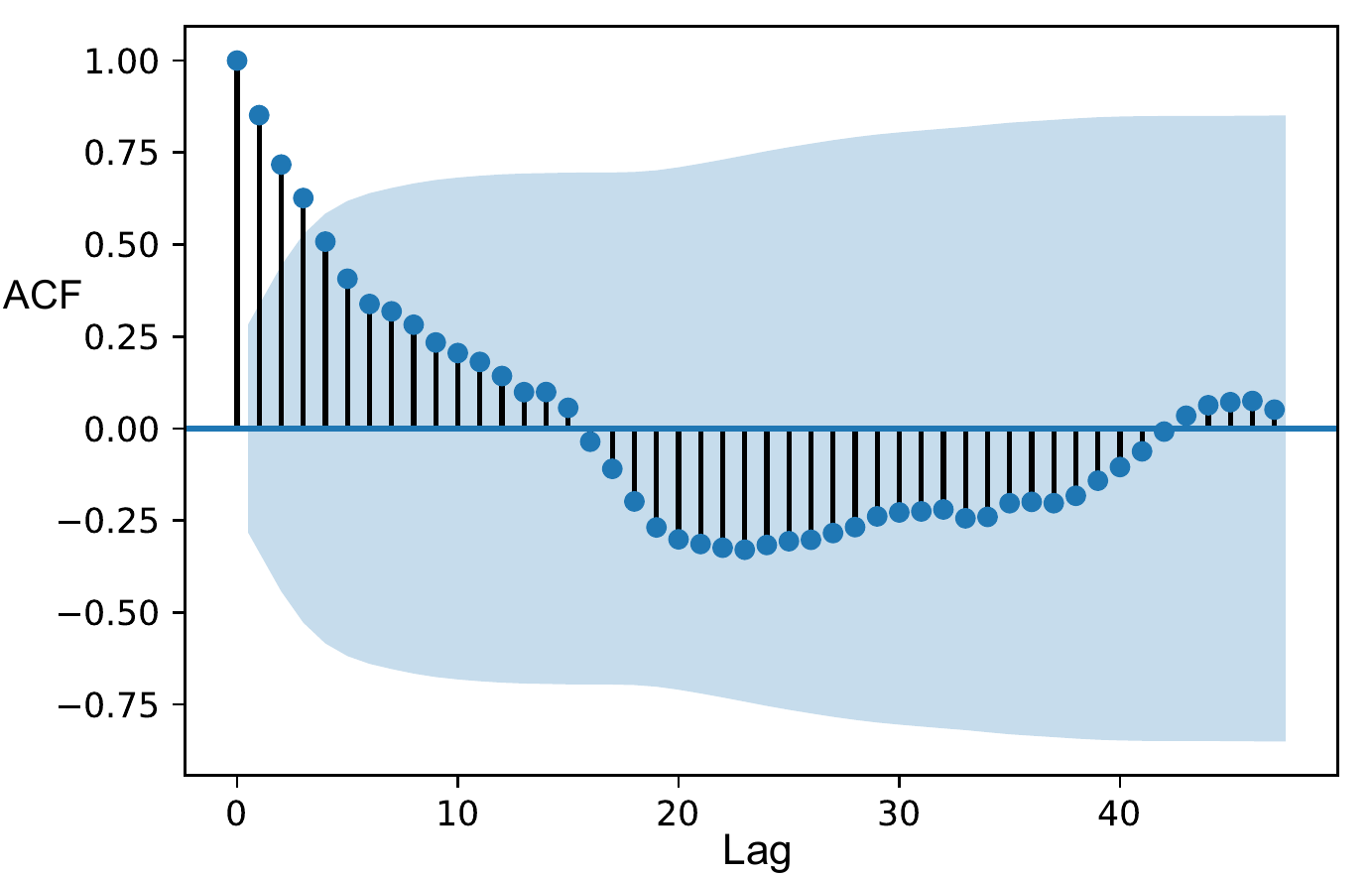}
\caption{The ACF of the SoC of one EV for several days from the dataset.}
\label{fig: one_day_four_acf_users}
\end{figure}

In Fig. \ref{fig: one_day_four_acf_users}, the auto correlation function (ACF) for one EV 
from the dataset is shown. The ACF is the similarity between observations (SoC values) as a function of the time lag between them. As shown in the figure, there is a positive correlation for the 
EV's ACF which is an indication of the regularity in the SoC charging patterns of a normal EV that can be used by our detector for detecting lying EVs.

\subsection{Malicious Dataset Creation}
Regarding the creation of malicious dataset, different attack scenarios are considered for reporting false SoC to the charging coordinator mechanism. These attacks are summarized in Table \ref{tab:Cyber Attacks}.
Four types of attacks are considered which are two versions of a partial reduction attack and two versions of selective time filtering attacks \cite{8746794}. 
The SoC of $EV_{i}$ at time $t$ and day $d$ is $S_{i}(d, t)$ and the reported SoC by $EV_{i}$ to the CC is $RS_{i}(d, t)$. If $EV_{i}$ is honest, then $RS_{i}(d,t)$ should be equal to $S_{i}(d,t)$ all the time.
As illustrated in Table \ref{tab:Cyber Attacks}, in \textit{Attack $\#1$}, the attacker tries to deceive the CC by multiplying the correct SoC value by a constant value $\alpha$ that is less than one to report lower SoC to gain high charging priority.
 In \textit{Attack $\#2$}, the EV tries to deceive the CC by manipulating the  SoC value that is sent to the CC by multiplying it by a value that is controlled by a time-dependent function $\beta_{i}(d, t)$ that should be less than one, e.g. $0.1\leq\beta_{i}(d,t)\leq0.8$. 
 In \textit{Attack $\#3$}, the EV uses a selective time filtering technique, where the malicious EV reports SoC of zero (or small value) during the interval \mbox{$t_{b}\leq t\leq t_{e}\\$,} otherwise, it reports its correct SoC $(S_{i}(d, t))$. \textit{Attack $\#4$} is similar to \textit{Attack $\#3$} but $EV_{i}$ selects a specific time interval \mbox{$t_{b}\leq t\leq t_{e}\\$ so} that the SoC values that are reported in this time interval are decreased gradually,
 and for other times, $EV_{i}$ reports its actual  SoC $(S_{i}(d, t)) $. In our dataset, the four attacks are applied on each row in the benign dataset producing a malicious dataset which contains $51,456$ records. 
{\renewcommand{\arraystretch}{2.5}
\begin{table}[t]
\centering
\caption{Attacks for reporting false SoC.}
\label{tab:Cyber Attacks}
\resizebox{1\columnwidth}{!}{%
\begin{tabular}{| 
>{\centering\arraybackslash}m{1.15cm}|
>{\centering\arraybackslash}m{1.85cm}| >{\centering\arraybackslash}m{5.5cm}|}
\hline 
Attack no. & Attack type & Formula\tabularnewline
\hline 
Attack \#1 & \multirow{2}{7em}{\hfill \vfill Partial reduction attack} & $RS_{i}(d,t)=\alpha S_{i}(d,t)$\\[7pt]
\cline{1-1} \cline{3-3}
Attack \#2 &  & $RS_{i}(d,t)=\beta_{i}(d,t)S_{i}(d,t)$\\[7pt]
\hline 
Attack \#3 & \multirow{2}{7em}{\hfill \vfill Selective time filtering attack} & $RS_{i}(d,t)=\begin{cases}
0&t_{b}\leq t\leq t_{e}\\
S_{i}(d,t) & else
\end{cases}$\\[7pt]
\cline{1-1} \cline{3-3}
Attack \#4 &  & $RS_{i}(d,t)=\begin{cases}
\beta_{i}(d,t) S_{i}(d, t_{b})&t_{b}\leq t\leq t_{e}\\
S_{i}(d,t) & else
\end{cases}$\\[7pt]
\hline 
\end{tabular}
}
\end{table}
}
 \subsection{Data Augmentation and Balancing}
\textit{Data augmentation} is the process of increasing the number of data samples for the purpose of creating an accurate and robust model \cite{Papernot:2017:PBA:3052973.3053009}. As it is known, as the number of samples increases in the dataset, the trained model becomes more accurate \cite{8729022}. \textit{Imbalance problem} occurs when one of the classes has more samples in the dataset than the other classes \cite{1549828}. 
Training a classifier model using imbalanced data could produce a classifier biased towards the class (classes) with the largest number of samples in the dataset. Therefore, data balancing techniques should be used when there are some data distributions that are significantly dominating other data distributions in the sample space to create accurate classifiers.
A solution for this problem is by using a data augmentation technique for increasing the number of samples in the minority class synthetically. The most common technique that is used for data augmentation for the purpose of balancing the dataset is the oversampling technique.  
Two common oversampling techniques \cite{ADASYN} are:
\begin{itemize}
    \item \textit{Sampling techniques:} In these techniques, the imbalanced dataset is compensated by oversampling, e.g., providing more copies of some of the samples of the minority class or undersampling the samples of the class with the majority number of samples. 
    \item \textit{Synthetic data generation:} In these techniques, the imbalance problem is solved by generating a synthetic data samples for balancing the original dataset. Common techniques that are under this category are the synthetic minority oversampling technique (SMOTE) \cite{Chawla02smote:synthetic} and the adaptive synthetic (ADASYN) sampling approach \cite{ADASYN}. 
\end{itemize}
In our dataset, we used the synthetic data generation technique because the sampling techniques suffer from the overfitting problem that could occur due to creating exact copies of the original data \cite{article}. 

\textit{ADASYN sampling approach} is a synthetic data generation technique for balancing minor classes in datasets and is considered an improved extension version of the SMOTE technique \cite{ADASYN}. In our dataset, the honest class is the minority that needs to be balanced with the malicious class. For using the ADASYN \cite{ADASYN} in our dataset, both classes should exist in the dataset, which are the minority class (honest class) and the majority class (malicious class). The reason for that is because ADASYN tries to balance the minority class with the majority class by calculating the ratio between the number of samples of the two classes for compensating the minority class with a larger number of synthetic samples to reach the number of the majority class samples. 
The ADASYN algorithm can be summarized in the following steps:
\begin{enumerate}
\item ADASYN computes the degree of the class imbalance in dataset: 
\[ratio=\frac{m_{min}}{m_{max}},\]\\
where, $m_{min}$ and $m_{max}$ are the number of samples in the minority and majority classes, respectively and that $ratio \in (0,1]$.
\item If $ratio<ratio_{th}$ where, $ratio_{th}$ is the present threshold for the maximum degree of tolerance for the class imbalance ratio, then:
\begin{itemize}
\item The total number of synthetic minority data samples need to be generated by ADASYN is :\[G=(m_{max}-m_{min})\xi,\] where, $\xi \in [0,1]$ is a parameter used to determine the required balance degree after finishing the process of synthetic data generation. When $\xi=1$, this indicates that a fully balanced dataset is required after the generation process is finished.
\item For each of the minority samples, ADASYN considers the k-nearest neighbors and calculates the $r$ value:\[r_{i}=\frac{\partial_{i}}{k},\] where $\partial_{i}$ is the number of samples in the $k$ nearest neighbors of sample $i$ based on the Euclidean distance rule in n-dimensional space that belongs to the majority class. $r_{i}$ value indicates the dominance of the majority class in the neighborhood. 
\item Before the number of synthetic samples to be generated is finally determined, ADASYN normalizes $r_{i}$ values for all minority samples so that their sum is one:\[\hat{r_{i}}=\frac{r_{i}}{\sum r_{i}},\] 
where \:\:\: $\sum \hat{r_{i}}=1.$\\
\item The ADASYN calculates the number of synthetic samples to be generated in each neighborhood as follows:\\
\[g_{i}=\hat{r_{i}}*G.\] 
Since $g_i$ is calculated using the respective $r_i$ value for every minority class sample, ADASYN creates more synthetic samples in neighborhoods with a greater ratio of majority to minority observations. 
\item For each sample $x_{i}$ from the minority class, there are $g_{i}$ samples to be generated by looping from $1$ to $g_{i}$ and doing the following:
\begin{itemize}
\item randomly choose a data sample $x_{j}$ from the $K$ nearest neighbors of the data sample $x_{i}$. 
\item generate synthetic observations:\\ \[s_{i}=x_{i}+(x_{j}-x_{i})\lambda,\]
where $s_{i}$ is the generated synthetic sample, $x_{i}$ is sample $i$ from the samples of the minority class, and $\lambda$ is a random number between $0$ and $1$.
\end{itemize}
\end{itemize}
\end{enumerate}
The ADASYN's main advantage  is that it uses a density distribution, $\hat{r_{i}}$, as a way to automatically determine the number of synthetic samples that should be generated for each sample in the minority class, 
which is considered as an advantage over the SMOTE algorithm which generates the same number of synthetic samples for each original minority sample. 
\\
\section{Proposed Lying Electric Vehicle Detector}
\label{Design of Electric vehicle detector}
In this section, we discuss the details of the proposed lying EV detector. 
The proposed detector is based on DNN classification and its designing process consists of two stages. 
The \textit{first} is a learning (training) stage, in which the structure and parameters of the detector are defined, and the \textit{second} is the stage of optimizing the classifier's hyper-parameters using a genetic optimization based approach for fine tuning. 
Two set of classifiers, namely, the GRU and MLP networks, are considered. 
\subsection{Multi-Layer Perceptron (MLP) Based detector}
As shown in Fig. \ref{fig: MLP}, the MLP detector consists of an input layer that receives the SoC values reported by an EV to classify it using nonlinear activation functions that exist in the middle and output layers. The network weights are updated in an iterative manner using two operations at each round. Firstly, given a set of training samples pairs in the form $(x_{i},y_{i})$, where $i=\{1, 2,\ldots, n\}$ and $n$ is the number of training samples used to train the network, the network output could be represented as $z=f(x,w)$, where $z$ is the output of the given input sample $x$, and $w$ is the weights-vector of the network. Secondly, the backpropagation process in which the gradient descent technique is used to update the network weights according to the error between the real label $y$ of the input sample and the output of the deep network $z$. 
Stochastic gradient descend (SGD) \cite{45428} is the algorithm which is used during the backpropagation phase for updating the weights of the network links. 

The weights are updated using the difference between the model output and the real label error according to Eq. \ref{eq:1}.
\begin{equation}
w^{t+1}\leftarrow w^{t}-\eta\nabla_{w}L(D^{t},w^{t})\label{eq:1}
\end{equation}
where $w^{t+1}$ is the weight vector of the network model at round
$t+1$, $w^{t}$ is the weight vector of the network model at the round
$t$, $\eta$ is the learning rate, $D^{t}$ is a subset of the training dataset $D$ at round $t$ (the mini batch), $L()$ is the error function that is used to compute the error between the real output and the predicted value (i.e., minimum square error), and $\nabla_{w}$ is the gradient of the error function or the differentiation of the error function with respect to weight $w$. The output layer of the MLP detector consists of two nodes to classify the input into one of two classes either to be honest or lying EVs.

\subsection{Gated Recurrent Unit (GRU) Based detector}
\begin{figure}[!t]
\centering
\captionsetup{justification=centering}
\includegraphics[width=2.4in]{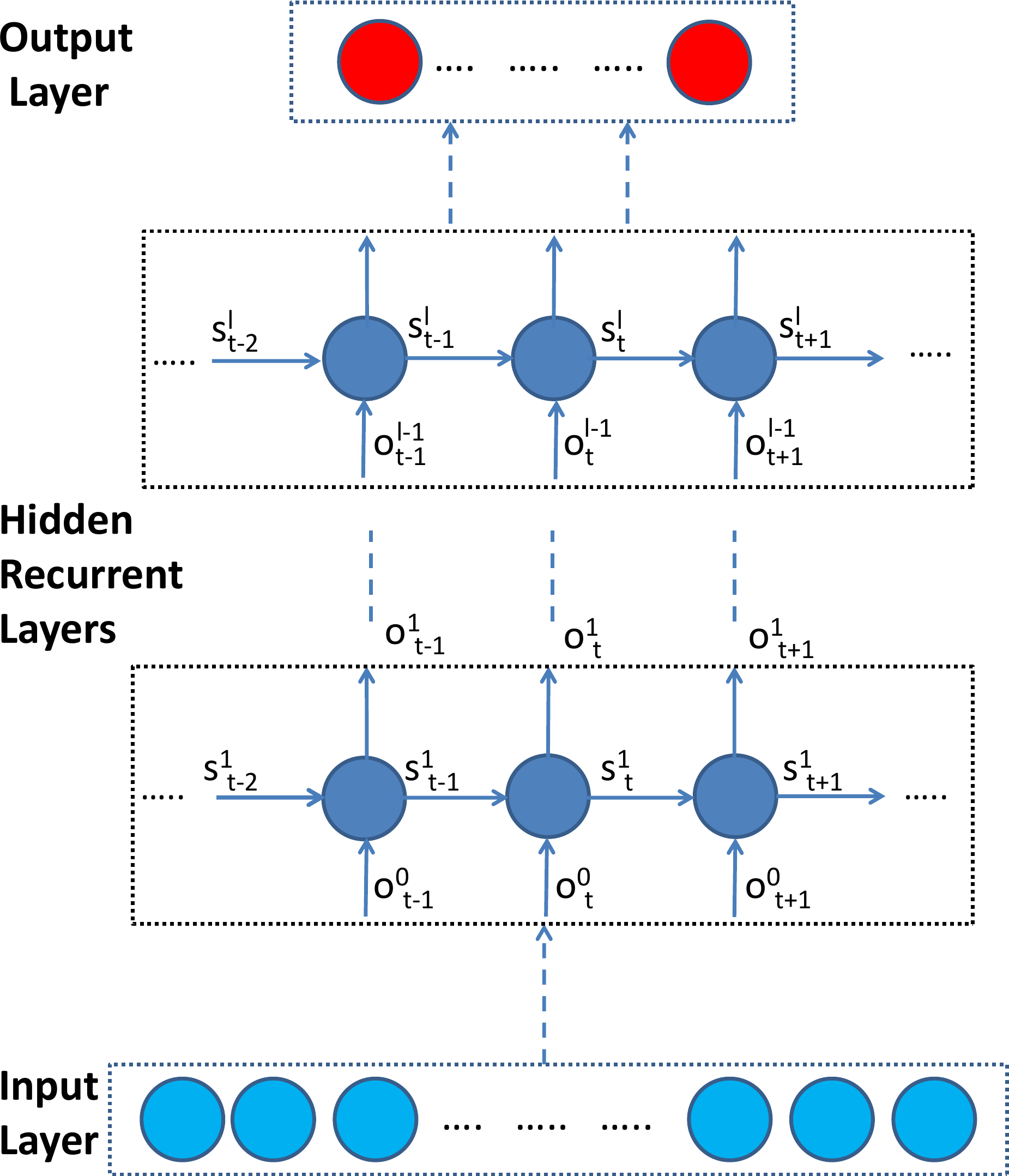}
\caption{Architecture of the gated recurrent unit (GRU) based detector.}
\label{fig: RNN_2}
\end{figure}

As shown in Fig.~\ref{fig: RNN_2}, the GRU detector consists of an input layer, a group of hidden layers and an output layer. The input layer is the one that is responsible for receiving the vector $RS_{i}(d,*)$ 
 that represents the SoC values of $EV_{i}$ during day $d$ at different reporting time periods $t \in \mathbb{T}$, where the vector has $T$ values. The detector has $L$ hidden layers following the input layer, where each consists of $N$ neurons with a given activation function. The input layer receives an input vector and sends data to middle layers that do computations and send data to output layer to produce an output vector, that has two neurons to classify a given input into either honest or liar. Therefore, the output of the last layer is two elements vector in the form $y(RS_{i}(d,*))=(0, 1)^{T}$ for a benign (honest) $EV_{i}$ or $y(RS_{i}(d,*))=(1, 0)^{T}$ in case of a lying $EV_{i}$. Given $L$ layers, for each layer $l \in \{1,2,\ldots,L\}$, the output is denoted by $o^l$ and $o^0=RS_{i}(d,*)$. Each hidden layer $l \in \{1,\ldots,L-1\}$, contains the following parameters:
\begin{itemize}
    \item The input at time $t$ is $o_{t}^{l-1}$, which is the output of the previous layer $l-1$.
    \item The hidden state $s_{t-1}^{l}$, that represents the memory computations using the previous layer's hidden state.
    \item The update gate of layer $l$ and state $t$ is computed using the input vector $o_{t}^{l-1}$ and previous hidden state $s_{t-1}^{l}$ as follows: $z_{t}^{l}=\sigma(o_{t}^{l-1}U_{z}^{l} + s_{t-1}^{l}W_{z}^{l} + b_{z}^{l})$. The learning process is responsible for the modification of the weight matrices $U_{z}^{l}$ and $W_{z}^{l}$ to reduce the error. $\sigma(\cdot)$ is the activation function while $b_{z}^{l}$ is the bias for the neurons in layer $l$. 
    \item 
    The \textit{reset} gate is calculated as follows: $r_{t}^{l}=\sigma(o_{t}^{l-1}U_{r}^{l}+s_{t-1}^{l}W_{r}^{l}+b_{r}^{l})$ where $U_{r}^{l}$, $W_{r}^{l}$, $U_{h}^{l}$, and $W_{h}^{l}$ are weights that are learned during the training process. 
    \item Finally, the current hidden state is computed as $s_{t}^{l}=(1-z_{t}^{l}) \odot h_{t}^{l} + z_{t}^{l}\odot s_{t-1}^{l}$, where $\odot$ is the Hadmard product and $h_{t}^{l}=tanh(o_{t}^{l-1}U_{h}^{l}+(s_{t-1}^{l} \odot r_{t}^{l})W_{h}^{l}+b_{h}^{l})$. The output at time $t$ is $o_{t}^{l}=softmax(V^{l}s_{t}^{l}+b_{o}^{l})$, where $V^{l}$ is a weight matrix that is learned by the training process.
\end{itemize}

\subsection{Hyper-parameters Optimization}
Optimum parameters improve the performance of the detectors. Although tuning the detectors' hyper-parameters is a difficult and time-consuming process because of the extensive computation required, an exhaustive search for hyperparameters is practically impossible, which motivates us to use the non-dominated sort genetic algorithm (NSGA-II) \cite{996017, 10.1007/3-540-45356-3_83} for finding an efficient solution in a reduced time compared with the exhaustive grid search \cite{350037}. 

In this paper, the NSGA-II is used to tune the hyper-parameters of the RNN architecture such as: the number of hidden layers $L$,  the number of neurons per layer $N$, the  type of optimization algorithm $O$, the initialization  method for the parameters $H$, the dropout rate $D$, the weight  constraint $J$, and the type of  activation functions at the  hidden layers $A_{hd}$ and output layers $A_{op}$. The NSGA is a multiple objective optimization algorithm and is an instance of evolutionary algorithms. 
There are two versions of the algorithm: the classical NSGA and the updated and currently canonical form, called NSGA-II. The NSGA's objective is to enhance the adaptive fit of a population of candidate solutions to a Pareto front restricted by a set of objective functions. The algorithm uses an evolutionary process with surrogates, including selection, genetic crossover and genetic mutation, for evolutionary operators. The population is sorted into a sub-population hierarchy based on the superiority of Pareto's order. An optimal choice of these  hyper-parameters significantly  improves the detection  performance. 

\section{Experiments and Results}
\label{exp}
In this section, the evaluation methodology that is used to measure the performance of the proposed detector is explained, and then the evaluation results are discussed. 
\subsection{Key Performance Metrics and Experiment Methodology}
We define the following key performance metrics which are used in our evaluation process: 
\begin{itemize}
\item \textsl{Detection Accuracy (ACC): } The ratio of the number of true positives and true negatives over the total number of EVs.
\begin{equation}
\textsl{Accuracy (ACC)}=\frac{TP+TN}{TP+TN+FP+FN}\label{eq:1-1}
\end{equation}
$TP$ represents true positives and is defined as the number of lying EVs that are correctly classified as lying. $TN$ is the true negatives and is defined as the number of honest EVs that are correctly classified as honest. $FP$ denotes false positives and is defined as the number of honest EVs incorrectly classified as lying. $FN$ represents false negatives and is defined as the number of lying EVs incorrectly classified as honest.
\\[2pt]

\item \textsl{True Positive Rate (TPR):} The ratio of the true positives to the total number of lying EVs 
\begin{equation}
\textsl{True Positive Rate (TPR)}=\frac{TP}{TP+FN}\label{eq:2}
\end{equation}

\item \textsl{False Positive Rate (FPR):} The ratio of the false positives to the total number of honest EVs 
\begin{equation}
\textsl{False Positive Rate (FPR)}=\frac{FP}{FP+TN}\label{eq:3}
\end{equation}




\item \textit{Detection Rate (DR):} The ratio of the number of truly detected lying EVs to the total number of EVs classified as lying.
\begin{equation}
\textsl{Detection Rate (DR)} =\frac{T_{P}}{T_{P}+F_{P}}\label{eq:10}
\end{equation}
\item \textit{False Acceptance Rate (FA):} It is the ratio of the number of honest EVs classified as lying to the number of honest EVs.
\begin{equation}
\textsl{False Acceptance (FA)} =\frac{F_{P}}{T_{N}+F_{P}}\label{eq:11}
\end{equation}

\item \textit{Highest Difference (HD):} it is the difference between detection rate (DR) and the false acceptance (FA).
\begin{equation}
    \textsl{Highest Difference (HD)} = DR - FA \label{eq:12}
\end{equation}

\item \textsl{Area Under Curve (AUC):} It is the area under the receiver operating curve (ROC). The ROC curve is a graphical way that captures the relationship between TPR and FPR \cite{NetworkRATDetection}. 

\end{itemize}

To evaluate the performance of the proposed detectors, the dataset is splitted into two parts. The first part constitutes 70\% of the whole dataset, and is used for the training process of the detectors. The second part is used in the test phase and constitutes about 30\% of the whole dataset. 
The GRU and MLP models are trained using the training dataset and evaluated using the testing dataset. The hyber-parameters of these models are adjusted using the non-dominated sort genetic algorithm (NSGA-II) to choose the best hyper-parameters that provide the best results. 

\subsection{Results and Discussion}
\begin{figure}[!t]
\centering
\captionsetup{justification=centering}
\includegraphics[width=\columnwidth]{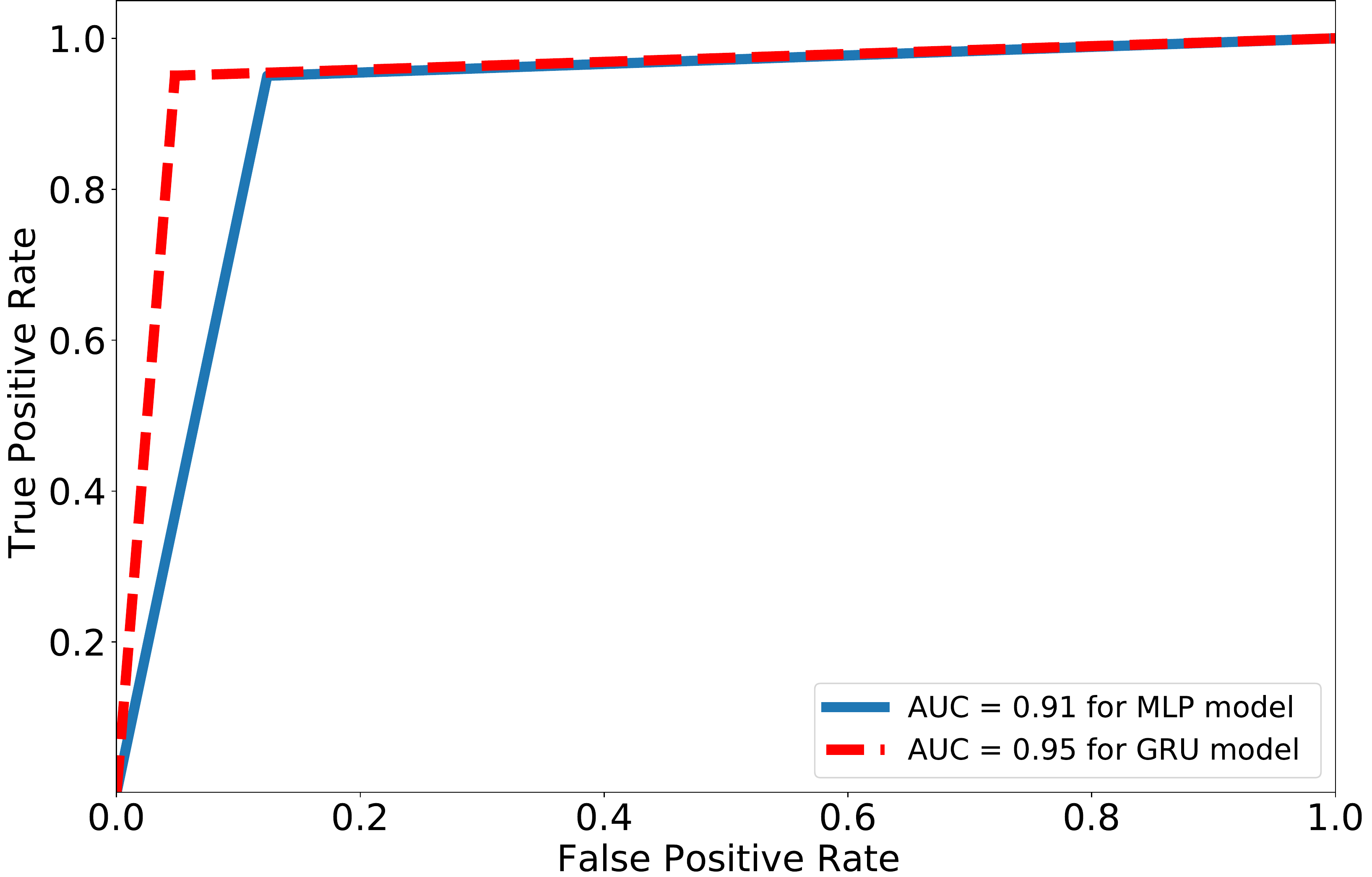}
\caption{The ROC curve for the GRU and MLP detector.}
\label{fig: RoC_curve}
\end{figure}

In this subsection, the results of the proposed detection models are discussed. In Table \ref{Final_Results_Table}, the results of the best network architectures, i.e., the architectures that provide the best results, for the GRU and MLP models are presented. 
In addition, the ROC curves of both architectures are presented in Fig. \ref{fig: RoC_curve}.

As it can be seen in Table \ref{Final_Results_Table}, the performance of the GRU detector outperforms that of the MLP detector with higher $DR$ and $ACC$ and lower $FA$. These results can be interpreted by the fact that the GRU has the ability to exploit the time series correlation in the SoC values that are reported from the EV for providing the best detection results. In addition, the good detection performance of the MLP comes on the expense of the high complexity of the architecture that is used to provide this performance.
The MLP's results given in Table \ref{Final_Results_Table} are provided using a complicated architecture that composes of six layers with $768$ neurons each with Softmax activation function at the output layer and ReLU activation function at the hidden layer.

On the contrary, the GRU detector provides the highest performance using a less complex architecture that composes of only two layers of $128$ neurons each, Softsign activation function at the hidden layer and Softmax activation function at the output layer, resulting in a much faster detector comparing to the MLP detector with higher performance. 
This reduction in the model complexity is also attributed to the nature of the GRU RNN that exploits the time series nature of the data providing good results with a much less complex model. The ROC curve in Fig. \ref{fig: RoC_curve} gives the performance of the two detectors in a visualization format using area under the curve metric which is $0.95$ for the GRU detector and $0.91$ for the MLP detector.  
{\renewcommand{\arraystretch}{1.35}
\begin{table}[t]
\centering
\caption{Evaluation results of deep neural network models.}
\label{Final_Results_Table}
\resizebox{1\columnwidth}{!}{%
\begin{tabular}{| >{\centering\arraybackslash}m{1.2cm}| >{\centering\arraybackslash}m{1.2cm}| >{\centering\arraybackslash}m{1.2cm}|
>{\centering\arraybackslash}m{1.2cm}|
>{\centering\arraybackslash}m{1.2cm}|}
\hline 
Classifier & DR & FA & HD & Accuracy \\\hline
GRU & 0.952 & 0.0495 & 0.903 & 0.951 \\\hline
MLP & 0.876 & 0.0497 & 0.827 & 0.913 \\\hline
\end{tabular}
}
\end{table}
}

\section{Conclusions}
\label{Conclusion}
Recently, EVs started to gain momentum due to their obvious advantages over gasoline vehicles. However, integrating EVs into the power grid requires a charging coordination mechanism to balance the charging load of the EVs with the power supply. The existing charging coordination mechanisms assume that the EVs report honest SoC values, although this is not beneficial to them. In this paper, we have first evaluated the impact of reporting false SoC values on the charging coordination mechanism. Then we have proposed a detector using deep learning techniques to identify lying EVs. Since the reported SoC values have a time series correlation, a deep neural network (DNN), has been selected to be the classifier that is used in the detection process. The selection process of the architecture of the deep network that provides the best performance is considered as an optimization problem with multi-objectives (detection rate and false acceptance rate). Accordingly, to solve this problem, we used the non-dominated sort genetic  algorithm (NSGA-II) to find the best architecture. 
However, for creating an accurate detector, a large dataset should be used in the training process. 
Therefore, we have created a new dataset and proposed a number of attacks and used them to create the malicious data.
The proposed detector has been evaluated and the results indicate that the detector achieves a high detection rate with a low false acceptance rate.

\section*{Acknowledgment}  

This work was supported by the Deanship of Scientific Research (DSR), King Abdulaziz University, Jeddah, under grant No. (DF-181-135-1441) The authors, therefore, gratefully acknowledge the DSR technical and financial support. 

\bibliographystyle{IEEEtran}
\bibliography{main} 

\end{document}